\documentclass[journal,twoside]{IEEEtran}
\usepackage{microtype}
\usepackage{amsmath,amsfonts,amssymb,graphicx}
\usepackage{cite}
\usepackage{physics}
\usepackage{multirow}
\usepackage[capitalize]{cleveref}
\usepackage{color}
\usepackage[dvipsnames]{xcolor}
\graphicspath{ {./FigureComponents/} }

\newlength{\singleColumnGraphWidth}
\newlength{\MatlabVerticalTighten}
\newlength{\extraVerticalSpace}
\newcommand{\normal}[2]{{\mathcal{N}(#1,#2)}}  
\newcommand{\iP}[1]{\mathrm{P}({#1})} 
\newcommand{\E}[1]{\mathrm{E}\!\left[\,{#1}\,\right]}   
\newcommand{\iE}[1]{\mathrm{E}[\,{#1}\,]}   
\newcommand{\variance}[1]{\mathrm{var}\!\left({#1}\right)}   
\def\smid{\,|\,}  
\def\sMid{\,;\,}  

\newcommand{\eqlabel}[1]{ \stackrel{(#1)}{=} }

\DeclareMathOperator*{\argmax}{arg\,max}
\DeclareMathOperator*{\median}{median}
\newcommand{\Frac}[2]{{{#1}/{#2}}}  

\newcommand{\openCase}  {\left\{ \begin{array}{@{\,}ll}}
\newcommand{\openCasell}{\left\{ \begin{array}{@{\,}ll}}
\newcommand{\openCasecl}{\left\{ \begin{array}{@{\,}cl}}
\newcommand{\openCaserl}{\left\{ \begin{array}{@{\,}rl}}
\newcommand{\openCaseTablell}{\left\{ \begin{array}{@{}ll}}
\newcommand{\openCaseTablecl}{\left\{ \begin{array}{@{}cl}}
\newcommand{\openCaseTablerl}{\left\{ \begin{array}{@{}rl}}
\newcommand{\closeCase} {\end{array} \right.}

\def\0{\mathbf 0}
\def\beamSigma{\sigma_{\mathrm b}}
\def\etaConv{\widehat{\eta}_{\mathrm{SM}}}
\def\etaClip{\widehat{\eta}_{\mathrm{clip}}}
\def\d{\mathsf{d}}
\def\CR{Cram{\'e}r--Rao}
\def\Poisson{\mathop{\mathrm{Poisson}}}
\def\ZTPoisson{\mathop{\mathrm{ZTPoisson}}}
\def\Polya{P{\'o}lya}

\def\binProb{b}
\def\binProbDiff{b'}
\def\mixMean{\zeta}
\def\numer{L} 
\def\gainPoisson{\beta_{\mathrm{Poisson}}}
\def\gainMixture{\beta_{\mathrm{mixture}}}

\def\mtilde{\widetilde{m}}
\def\Mtilde{\widetilde{M}}
\def\Ttilde{\widetilde{T}}
\def\Xtilde{\widetilde{X}}
\def\xtilde{\widetilde{x}}
\def\etaBaseline{\widehat{\eta}_{\rm baseline}}
\def\Imixture{\mathcal{I}^{\mathrm{mix}}}
\def\IPoisson{\mathcal{I}^{\mathrm{Poisson}}}
\def\ITRM{\mathcal{I}_{\mathrm{TRM}}}
\def\ITRMmixture{\Imixture_{\mathrm{TRM}}}
\def\ITRMPoisson{\IPoisson_{\mathrm{TRM}}}
\def\Iscan{\mathcal{I}_{\mathrm{scan}}}
\def\Iscanmixture{\Imixture_{\mathrm{scan}}}

\def\Lscanmixture{\mathcal{L}_{\mathrm{scan}}^{\mathrm{mix}}}

\def\LscanPoisson{\mathcal{L}_{\mathrm{scan}}^{\mathrm{Poisson}}}

\def\gammaMLE{\widehat{\gamma}_{\mathrm{MLE}}}
\def\gammaMMLE{\widehat{\gamma}_{\mathrm{MMLE}}}
\def\YieldPotential{\varphi}

\title{Beam Cross Sections Create Mixtures: Improving Feature Localization in Secondary Electron Imaging}

\author{Vaibhav Choudhary,
Akshay Agarwal, and Vivek K Goyal
\thanks{V. Choudhary and V. K. Goyal are with the Department of Electrical and Computer Engineering, Boston University, Boston, MA 02215 USA (e-mail: vchoudh@bu.edu; v.goyal@ieee.org).}
\thanks{A. Agarwal was with Boston University at the initiation of this work and is now with the Department of Physics, Kennesaw State University, Kennesaw, GA 30144 USA (e-mail: akshayagarwal019@gmail.com).}
\thanks{This work was supported in part by a gift from Dr.\ John Zheng Sun and in part by a 2024 Guggenheim Fellowship.}
}

\begin{document}
\maketitle

\begin{abstract}
Secondary electron (SE) imaging techniques, such as scanning electron microscopy
and helium ion microscopy (HIM),
use electrons emitted by a  sample
in response to a focused beam of charged particles incident at a grid of raster scan positions.      
Spot size---the diameter of the incident beam's spatial profile---is one of the limiting factors for resolution,
along with various sources of noise in the SE signal.
The effect of the beam spatial profile is commonly understood as convolutional.
We show that under a simple and plausible physical abstraction for the beam,
though convolution describes the mean of the SE counts,
the full distribution of
SE counts is a mixture.
We demonstrate that this
more detailed modeling can enable resolution improvements over conventional estimators through a stylized application
inspired by
semiconductor inspection: localizing the edge in a two-valued sample.
We derive Fisher information about edge location in conventional and time-resolved measurements (TRM)
and also derive the maximum likelihood estimate (MLE) from the latter.
Empirically, the MLE computed from TRM is approximately efficient except at very low beam diameter,
so Fisher information comparisons are predictive of performance and can be used to optimize the beam diameter relative to the raster scan spacing.
Monte Carlo simulations provide an example of the MLE giving a 5-fold reduction in
root mean-squared error (RMSE) of edge localization
as compared to conventional interpolation-based estimation.
The RMSE is substantially below both the beam diameter and the raster scan spacing and thus sub-pixel localization is demonstrated.
Applied to three real HIM datasets,
the average RMSE reduction factor is 5.4.
\end{abstract}

\begin{IEEEkeywords}
electron beams,
electron microscopy,
Fisher information,
focused ion beams,
mixture models,
particle beam microscopy,
Poisson compounding,
semiconductor inspection,
spot size effect,
sub-pixel localization, 
zero-truncated Poisson mixture.
\end{IEEEkeywords}

\section{Introduction}
\label{sec:introduction}

\IEEEPARstart{N}{anometer}-scale imaging is central to discovery and everyday diagnostics in a variety of scientific and engineering fields.
It is essential in semiconductor manufacturing,
for example,
where the shrinking in device sizes over the last 50 years
would not have been possible without precise measurements
of each generation of devices~\cite{Orji2018}.
In biology, nanoscale imaging enables scientists to observe
structures and processes at the subcellular level~\cite{Dawson2021, Quist2012}.
Optical techniques reach nanoscale for specific samples with fluorescent dyes and fluorescent genetic reporters~\cite{Wessels2010};
for general samples, however, diffraction precludes such fine resolution.
Instead, the most prevalent instruments for nanoscale imaging detect secondary electrons (SEs)
generated by a highly focused incident beam of charged particles,
raster scanned on a 2D grid~\cite{Reimer1998}.

The micrographs produced by SE imaging instruments 
show spatial variation of \emph{SE yield}---defined as the mean number of SEs emitted
by the sample per incident particle---subject
to various sources of noise.
Though the dependence of SE yield on local sample properties is complicated,
SE imaging is nevertheless ubiquitous in semiconductor metrology,
where it is used for
critical dimension
measurements~\cite{Villarrubia2015},
line width roughness
analysis~\cite{Villarrubia2005, Azarnouche2012, Hiraiwa2010, Constantoudis2013, Verduin2014, Constantoudis2018},
characterizing cross-sectional samples~\cite{Krueger1999, Young2005},
and analyzing and classifying sample defects~\cite{Postek1987, Nakagaki2009, Nakamae2021}.
Biological applications of SE imaging have been abundant for a half century~\cite{HaynesPease1968}
and are especially popular when combined with focused ion beam (FIB)
sectioning for volumetric imaging~\cite{Narayan2015, Kizilyaprak2014, Titze2016}.

The noise in SE images comes from three mechanisms~\cite{Sim2004,Ura1993}:
\emph{source shot noise}, variation in the number of incident particles in a
fixed time period;
\emph{target shot noise}, variation in the number of emitted SEs
for a single incident particle; and
\emph{detector noise}, variation in measured response to a number of SEs
(including the efficiency in emitted SEs reaching the detector).
Additionally, the spot size or cross-sectional profile of the incident particle beam
is not generally considered a source of noise but it too influences the image quality.

Despite the widespread use of SE imaging,
works that examine the information content in SE detector signals
and seek statistically optimal estimation of SE yield
have appeared only recently.
Peng et al.~\cite{PengMBBG:20, PengMBG:21} introduced the concept of \emph{time-resolved measurement} (TRM)
to SE imaging under an assumption of ideal SE detection.
TRM can be understood to almost completely mitigate source shot noise; see \cref{sec:TRM}.
Agarwal et al.~\cite{AgarwalPG:23} introduced modeling of
actual
detector voltage signals
and experimentally demonstrated
noise reduction in a helium ion microscope (HIM)~\cite{Agarwal:2024-PNAS}.
These previous works do not include any modeling of beam cross section.

In this paper, we complement the framework of previous works with
detailed physics-based probabilistic modeling of the
beam cross section.
Our main assertion is that nonzero beam cross-sectional area causes the SE count
due to a single incident particle to follow a mixture distribution.
This contrasts with a convolutional model that may be intuitive because,
na\"ively, the effect of the beam seems analogous to blurring in a linear optical system.
Indeed, convolutional modeling is conventional in scanning electron microscopy~\cite{Frase2007, Mack2015, Villarrubia2015};
we seek to expose the limitations and suboptimality of this modeling
while providing an operational alternative.

As a simplified abstraction of a prototypical problem in semiconductor metrology,
we consider here the localization of an axis-aligned straight edge
in a two-dimensional sample
with distinct SE yield levels on each side of the edge.
This simplified imaging-based inspection problem allows us to
use root mean-squared error (RMSE)
of the edge location
to explicitly
compare the accuracies of
estimates produced from mixture modeling and from convolutional modeling.
In a highlighted example
(see \cref{subfig:edge_50.2_dose_eta_1_10_RMSE}(b)),
Monte Carlo simulations
show RMSE reduced by a factor of 2.5 by using TRM with a mixture model as compared to TRM with a convolution model;
improvement relative to a conventional interpolation method is by a factor of 5\@.
In general, improvements will depend on many factors,
including beam diameter relative to raster scan spacing,
edge location relative to the scan grid,
SE yield values,
and number of incident particles.
On experimental HIM data, we observe an average RMSE reduction factor of 1.5 with respect to a convolution model and 5.4 with respect to interpolation.

Advantages from our novel mixture modeling are small in the absence of TRM,
which may explain why this type of modeling has not been developed previously.
Here, this dependency necessitates detailed discussion of spatial and temporal beam characteristics
and the properties of TRM itself
before we turn to the application of edge localization.
\Cref{sec:setup} introduces the models of the beam (\cref{subsec:beam_sample_models}),
the sample (\cref{subsubsec:sample-char}),
and their interaction (\cref{subsec:se_distribution_one_ion})
that underlie this work.
We establish there a physical justification for a two-component Poisson mixture model for SE counts
when the sample is two-valued
with SE yields $\eta_1$ and $\eta_2$
(\cref{sec:two-valued}).
Though we are not yet considering the edge localization problem \emph{per se}---and for clarity we do not include source shot noise---the
fundamental gap between the best uses of TRM and non-TRM data when
$\eta_1$ and $\eta_2$ are known (\cref{subsec:FI_deterministic})
motivates our entire study.
The challenge is to realize the potential gains from a mixture model, translated to edge localization, in the face of source shot noise (\cref{sec:challenges}).
Analyses in \cref{sec:setup} are for observations of SE counts due to a single incident particle and for repetitions of such hypothetical observations for a deterministic number of trials.
In practice, particle incidences are random and not perfectly inferrable.
Mitigating this effect was the motivation for time-resolved measurement.
\Cref{sec:TRM} reviews---and in some cases slightly generalizes---results from~\cite{PengMBG:21,AgarwalPG:23}
on the advantages from TRM when beam cross-sectional effects are neglected.
Measurement models are formally introduced (\cref{subsec:measurement-models}),
and analyses are provided for conventional (\cref{sec:conventional-analyses})
and time-resolved measurements (\cref{sec:Poisson-TRM-analyses}).
\Cref{sec:truncated-mix} brings together the spatial (\cref{sec:setup}) and temporal (\cref{sec:TRM})
aspects of our incident beam model
for the estimation of two-component Poisson mixtures,
for conventional (\cref{subsec:analysis-conventional})
and time-resolved measurements (\cref{sec:TRM-mixture-analyses}).
Using the relationship between edge location, mixture fraction, and SE yield,
\cref{sec:edge-problem}
analyzes estimation of edge location
when combining data acquired from a line of scan locations across an edge.
We show that accuracy of localization can be finer than scan grid spacing,
thereby achieving sub-pixel resolution. 
We furthermore establish a method for optimal selection of beam width.
In Monte Carlo simulations of estimator performance,
we observe that the mixture estimator has lower mean-squared error (MSE)
than other estimators of the edge location.
The same performance trends are also observed on experimental datasets acquired on a helium ion microscope.

\section{Underlying Models and Problem Setup}
\label{sec:setup}

\begin{figure}
  \centering
  \begin{tabular}{cc}
  \multirow{3}{0.3\linewidth}[0.16\linewidth]{\includegraphics[width=\linewidth]{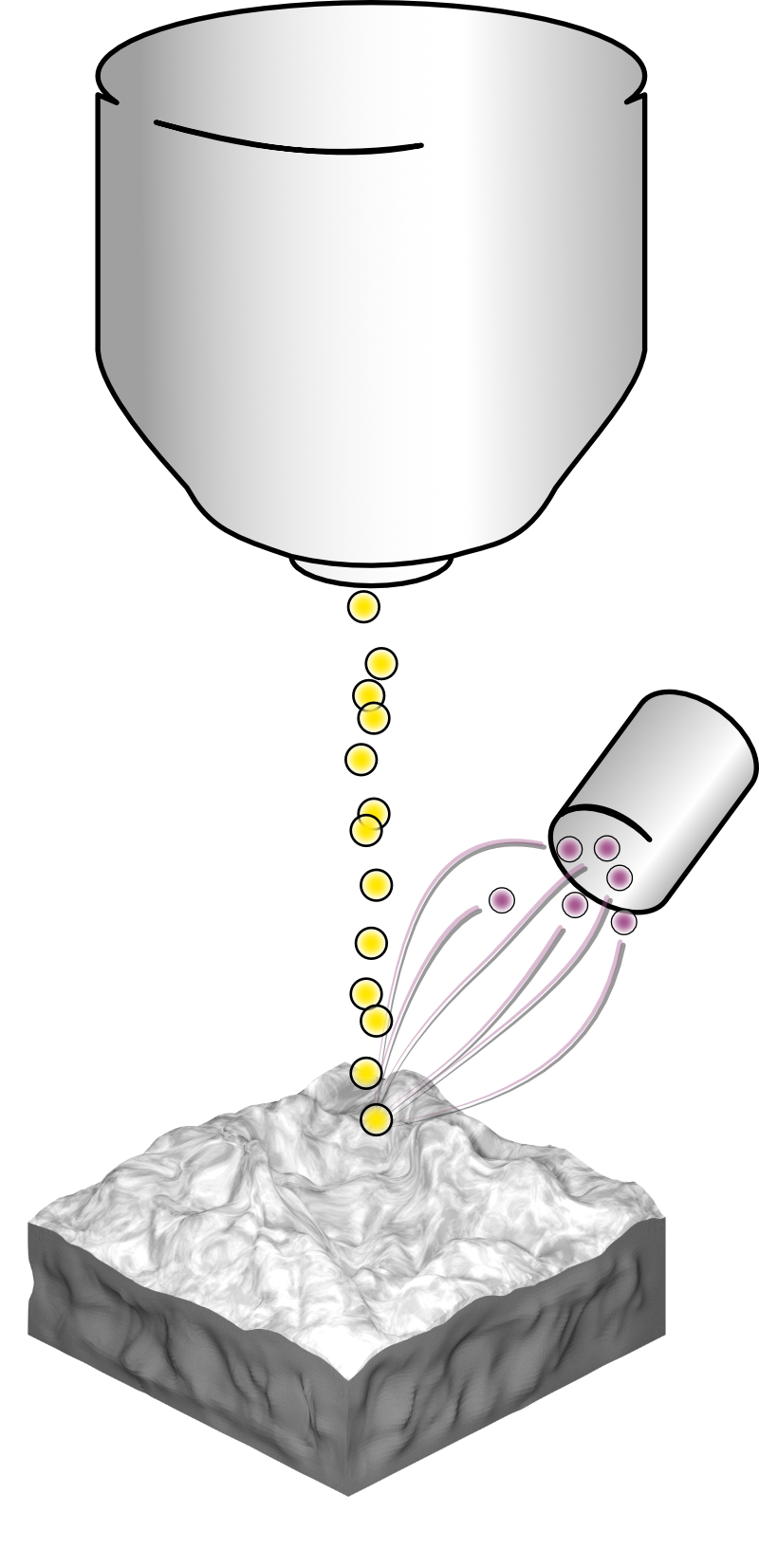}}
  & \includegraphics[width=0.6\linewidth]{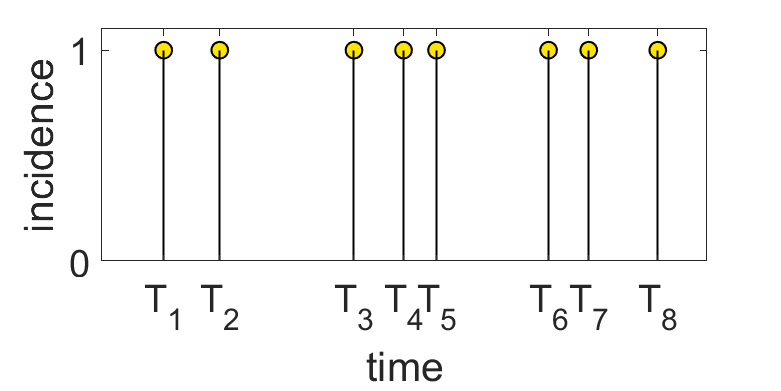}
  \\
  & {\footnotesize (b) Temporal distribution} \\
  & \includegraphics[width=0.6\linewidth]{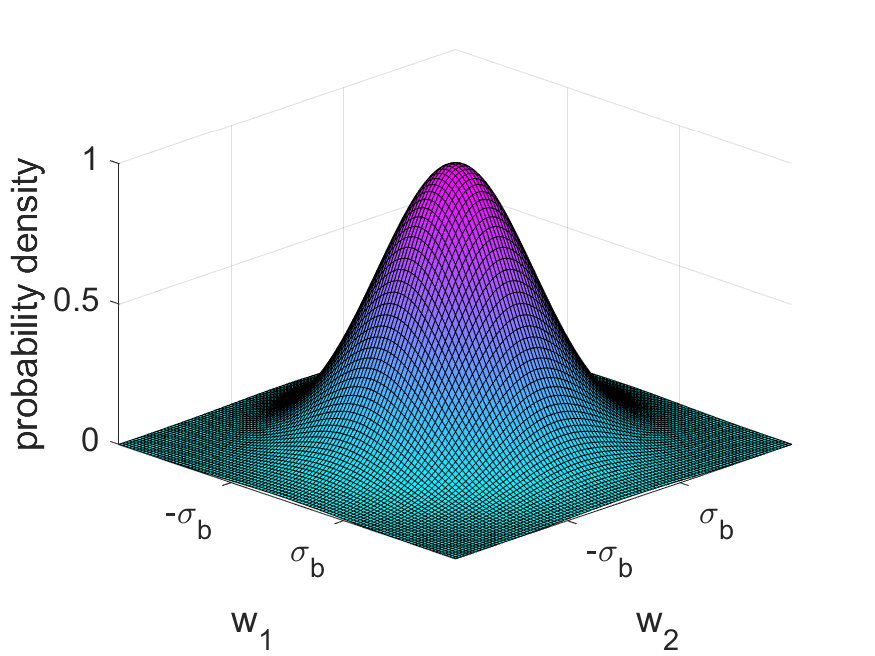}
  \\
  {\footnotesize (a) Schematic} &
  {\footnotesize (c) Transverse spatial distribution}
  \end{tabular}
\caption{Model of data collection.
(a) Depiction of incident particle beam and generation of SEs.
(b) Temporally, incident particles arrive as a Poisson process (one representative realization shown). 
(c) Spatially, deviations from the nominal raster scan location follow a Gaussian distribution.
}
\label{fig:-pbmmodel}
\end{figure}

SE imaging modalities
involve data collection at a grid of
raster scanned positions of
a focused beam of particles. 
The incident beam may consist of electrons or
ions, depending on the type of microscope;
we will sometimes call these particles ions for brevity.
Electrons emitted by
the sample
(contrasted with backscattered particles)
are called secondary electrons
and are detected without directional resolution
because they are deflected by an electrostatic field,
as illustrated in \cref{fig:-pbmmodel}(a).

The detector is typically of the Everhart--Thornley (ET) type,
consisting of a scintillator followed by a light pipe--photomultiplier couple~\cite{Everhart1960, Reimer1998, WardNE:06, WardNE:07}.
In all contemporary instruments,
the ET detector voltage signal is integrated
(or summed digitally after analog-to-digital conversion)
over a scan dwell time to produce a pixel value.
Here, as in recent research that has not yet reached widespread practice~\cite{PengMBBG:20,PengMBG:21,Seidel2022TCI,PengKSYG:23,Agarwal:2024-PNAS},
we consider more sophisticated processing of detector signals to improve imaging capabilities.
To avoid obscuring the main contributions of the paper,
we will omit consideration of the detector noise introduced by the ET detector,
instead assuming that a detector accurately and instantaneously counts SEs.
When presenting results based on experimental data in \cref{sec:experimental},
we will discuss the hard-decision estimation approach that is applied to ET detector voltage signals to obtain proof-of-principle results.

\subsection{Model of Incident Particle Beam}
\label{subsec:beam_sample_models}

As shown in the cartoon of \cref{fig:-pbmmodel}(a),
incident particles are haphazard in both transverse spatial positions and times of incidence.
We model the beam temporally as arrivals in
a Poisson process with a known rate $\Lambda$ per unit time (see \cref{fig:-pbmmodel}(b)).
Let $t$ denote the pixel dwell time.
Then the number of particles $M$ incident on a pixel is a Poisson random variable with parameter $\lambda=\Lambda t$.
We refer to $\lambda$ as the \emph{dose};
normalization by beam cross-sectional area or raster scan density is helpful when assessing likelihood of sample damage.
The value of $M$ is not directly observable; its variation from $\iE{M} = \lambda$ is called \emph{source shot noise}.

The position of an incident particle
deviates
from the nominal raster scan position.
We model this with a circularly symmetric 2D Gaussian distribution:
$\mathbf{W} = (W_1,W_2) \sim \normal{\0}{\beamSigma^2 I_2}$
(see \cref{fig:-pbmmodel}(c)).
We will refer to the beam standard deviation $\beamSigma$ as the \emph{beam width};
we could instead use an equivalent specification of full width at half maximum (FWHM).

In an ion- or electron-beam instrument, the values of beam intensity $\Lambda$ and beam standard deviation $\beamSigma$ depend on many factors and
are controllable to some extent, with limitations that depend on the particle type.
Increasing $\Lambda$ is generally associated with increasing $\beamSigma$.
Increasing
working distance is also associated with higher $\beamSigma$.
Typical minimum values of $\beamSigma$ are in the subnanometer range,
but larger values are chosen (e.g., through beam defocus) to limit sample damage.
One contribution of this paper is to study favorable ratios between $\beamSigma$ and raster scan spacing.

\subsection{Model of Sample}
\label{subsubsec:sample-char}
As noted in the introduction,
SE imaging generally shows the variation in SE yield.
To work toward a precise understanding of the effect of the beam cross-section,
we make a distinction between the SE yield at a scan position
and the underlying SE yield of the sample at an incidence location.
For this, we require a continuous-space description of the sample even though data is collected by discrete raster scanning of the beam.

We normalize the transverse spatial dimensions of the sample by
the imaging raster scan spacing so that
raster scan locations are on the grid $\{0,\,1,\,\ldots,\,\ell-1\}^2$.
Our mathematical model of the sample
is not restricted to the raster scan grid
because we wish to model effects of the beam cross section and
the potential for sub-pixel resolution.
Thus, the sample is defined as a function
$\YieldPotential : {\mathbb{R}}^2 \rightarrow [0,\infty)$
that maps from the ion incidence position
(which, due to nonzero beam width, is not exactly a raster scan position)
to the SE yield.
 
When an ion is incident at $\mathbf{s} = (s_1,s_2)$, the number of emitted SEs is not deterministically
$\YieldPotential(\mathbf{s})$.
Distinct mechanisms of emission have been established.
Potential emission depends mostly on incident particle charge
and results in SE counts that can be modeled as binomial~\cite{VanaALW:95, Eder1997};
it is especially significant with multiply-charged projectiles~\cite{Schwestka2019, Niggas2022}.
Kinetic emission depends mostly on incident particle mass and velocity
and results in SE counts that can be modeled as Poisson when
cascade effects are minimal and {\Polya} more generally~\cite{Benka1995}.
Due to the combination of mechanisms and the crudeness of the theoretical models,
no single parametric model fits measured data well across a wide range of conditions.
Many studies of SE emission and all previous SE imaging methods,
to the best of our knowledge,
treat the number of SEs as a Poisson random variable, including effects of nonzero beam width.
Here, we assume the SE counts are Poisson distributed when effects of nonzero beam width are \emph{not} included. 

The problem of SE imaging in general is the estimation of the function
$\YieldPotential$,
usually on a subset of the domain such as $[0,\,\ell-1]^2$
or only on the scan grid $\{0,\,1,\,\ldots,\,\ell-1\}^2$.
In \cref{sec:two-valued},
we will restrict
$\YieldPotential$
to model a two-valued sample with a single axis-aligned edge.

\subsection{Model of SE Distribution for One Incident Particle}
\label{subsec:se_distribution_one_ion}

Under the beam and sample models in \cref{subsec:beam_sample_models,subsubsec:sample-char},
one incident ion at grid scan location $\mathbf{g} = (g_1,g_2)$
will actually strike the sample at $\mathbf{s} = \mathbf{g} + \mathbf{W}$,
where $\mathbf{W} \sim \normal{\0}{\beamSigma^2 I_2}$.
Since the sample has SE yield
$\YieldPotential(\mathbf{s})$
for this incidence position,
the random number of generated SEs is
\begin{equation}
    \label{eq:X-distribution}
    X \sim \Poisson( \YieldPotential( \mathbf{g} + \mathbf{W} ) ).
\end{equation}
Since the random quantity
$\YieldPotential( \mathbf{g} + \mathbf{W} )$
is the parameter of a Poisson random variable,
this $X$ has
a \emph{Poisson mixture}
distribution~\cite{Karlis2005}.

By iterated expectation, 
the mean of $X$ is
\begin{align}
    \E{X} &= \E{ \E{ X \smid \mathbf{W} }} 
       = \E{ \YieldPotential(\mathbf{g} + \mathbf{W}) } \nonumber \\
      &= \int_{-\infty}^{\infty} 
         \int_{-\infty}^{\infty} 
         \YieldPotential(\mathbf{g} + \mathbf{w}) f_{\mathbf{W}}(w_1,w_2) \, \d w_1 \, \d w_2.
         \label{eq:mean-X}
\end{align}
We use $\eta$ as a compact shorthand for $\E{X}$,
which is the SE yield incorporating the beam model,
in contrast to the idealized $\YieldPotential(\mathbf{g})$.
The convolutional form in \eqref{eq:mean-X} is reminiscent of a linear blur kernel in an optical system.
It is, however, central to our premise to describe the full distribution
rather than merely the mean.

A simple way to illustrate that
$X$ is not Poisson distributed---and hence the effect of the beam cross-section is not merely a convolutional blurring of
$\YieldPotential(\mathbf{s})$,
decoupled from the distribution family of the target shot noise---is to compute its variance.
Using the law of total variance:
\begin{align}
    \variance{X}
      &= \E{ \variance{ X \smid \mathbf{W} } } + \variance{ \E{ X \smid \mathbf{W} } } \nonumber \\
      &= \E{ \YieldPotential( \mathbf{g} + \mathbf{W} ) } + \variance{ \YieldPotential( \mathbf{g} + \mathbf{W} ) }.
  \label{eq:var-X}
\end{align}
Since Poisson-distributed $X$ would have variance equal to its mean,
we will refer to
\begin{equation}
  \label{eq:excess-variance}
    \variance{X} - \E{X}
    = \variance{ \YieldPotential( \mathbf{g} + \mathbf{W} ) }
\end{equation}
as the \emph{excess variance} of the mixture model.
It
is nonnegative,
and it is zero only in the limit of
$\YieldPotential$
having no variation
in the neighborhood of the nominal scan location $\mathbf{g}$, where the relevant neighborhood size depends on $\beamSigma$.

\subsection{Samples with Two SE Yield Values}
\label{sec:two-valued}
\begin{figure}
 \centering
  \begin{tabular}{@{}c@{\,\,\,}c@{}}
  \includegraphics[width=0.36\linewidth]{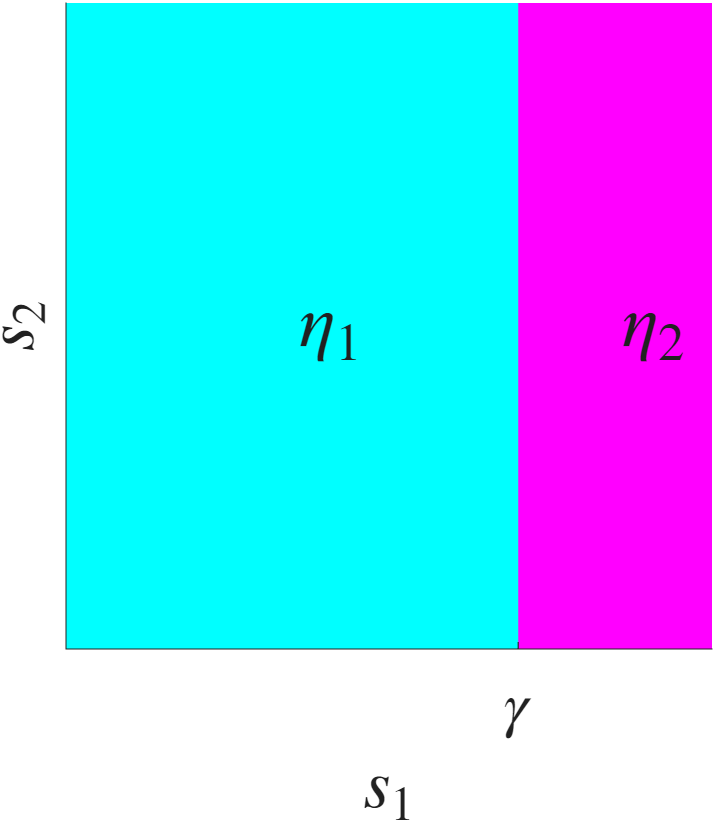} &
  \includegraphics[width=0.62\linewidth]{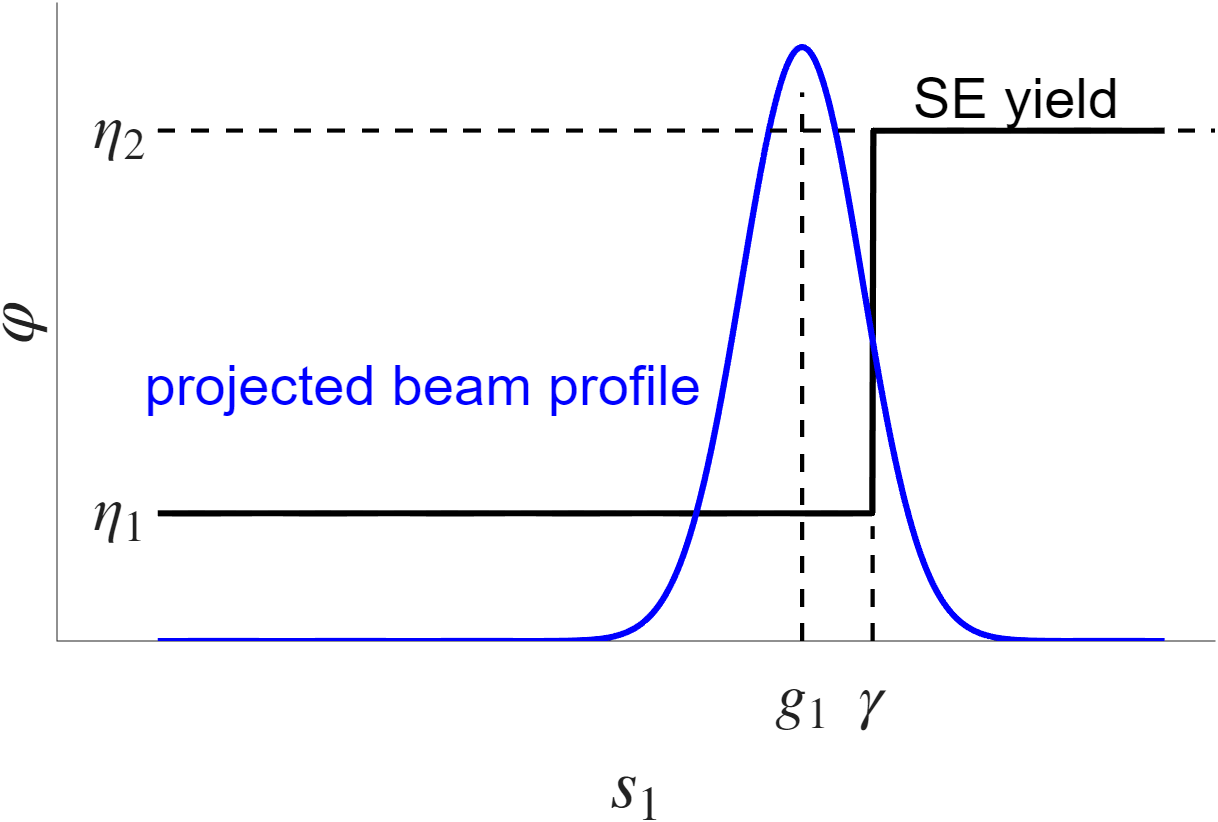}
  \\
  {\footnotesize (a) Two-valued sample} &
  {\footnotesize (b) Projections to 1D functions}
  \end{tabular}
\caption{A sample with two SE yield values separated by a vertical edge at horizontal position $\gamma$.
(a) Depiction of the function
$\YieldPotential(s_1,s_2)$.
(b) Sample and beam distribution projected to 1D functions of only the horizontal position.
}
\label{fig:twoValuedSample}
\end{figure}

As a prototype for a semiconductor sample,
consider
\begin{equation}
    \label{eq:two-valued}
    \YieldPotential(s_1,s_2) = \openCaserl
       \eta_1, & s_1 \leq \gamma; \\
       \eta_2, & s_1 >  \gamma,
       \closeCase
\end{equation}
as illustrated in \cref{fig:twoValuedSample}(a).
A step in semiconductor inspection or manufacturing process characterization could include imaging of this sample.
Except when discussing the value of \emph{a priori} knowledge of $\eta_1$ and $\eta_2$ in \cref{subsec:FI_deterministic}, we will assume $\eta_1$ and $\eta_2$ are known,
with $\eta_1 < \eta_2$.
The imaging task then reduces to estimation of edge position $\gamma$.

Since the sample has no vertical variation and
the Gaussian beam model in \cref{subsec:beam_sample_models} has independent horizontal and vertical deviations
from the raster scan position,
the number of generated SEs $X$ depends only on the horizontal raster scan position.
Specifically, from \eqref{eq:X-distribution},
the distribution of
$\YieldPotential(\mathbf{g}+\mathbf{W})$
determines the distribution of $X$.
Since \eqref{eq:two-valued} has no dependence on $s_2$
and $W_1$ is independent of $W_2$,
we are left with projections to 1D functions as shown in \cref{fig:twoValuedSample}(b).
The projected beam profile is Gaussian with standard deviation $\beamSigma$.

The two-valued nature of
$\YieldPotential$
allows us to simplify further.
Define
\begin{align}
    q =&           \ \iP{\YieldPotential(\mathbf{g}+\mathbf{W}) = \eta_2} \nonumber \\
      \eqlabel{a}& \ \iP{g_1+W_1 > \gamma} 
      \eqlabel{b} \ 1 - \Phi( (\gamma - g_1)/\beamSigma ), 
      \label{eq:mixing-q}
\end{align}
where (a) follows from \eqref{eq:two-valued}; and
(b) uses $\Phi(\cdot)$ to denote the standard normal cumulative distribution function.
Using this $q$ in \eqref{eq:X-distribution} makes $X$ a two-component Poisson mixture:
\begin{equation}
    \label{eq:X-distribution-two-component}
    X \sim
      \openCaserl
    \Poisson( \eta_1 ), & \mbox{with probability $1-q$}; \\
    \Poisson( \eta_2 ), & \mbox{with probability $q$},
       \closeCase
\end{equation}
meaning that the probability mass function (PMF) of $X$ is
\begin{equation}
    \label{eq:mixpoissondistr}
    \mathrm{P}_X(x \sMid q, \eta_1, \eta_2) = 
    (1-q)\frac{\eta_1^{x}e^{-\eta_1}}{x!} +
    q\frac{\eta_2^{x}e^{-\eta_2}}{x!} .
\end{equation}
Since $\Phi$ is one-to-one, an estimate of $q$ can be mapped to an estimate of edge position $\gamma$ using \eqref{eq:mixing-q}.

Using the law of iterated expectation, the mean of $X$ is
\begin{equation}
    \label{eq:mixture-mean-two-component}
    \E{X} = (1-q)\eta_1 + q\eta_2
\end{equation}
and, using the law of total variance, the variance of $X$ is
\begin{equation}
    \label{eq:mixture-variance-two-component}
    \variance{X} = (1-q)\eta_1 + q\eta_2 + q(1-q)(\eta_2 - \eta_1)^2.
\end{equation}
The excess variance---variance not accounted for by the convolutional model---is
\begin{equation}
  \label{eq:excess-variance-two-valued}
    \variance{X} - \E{X}
    = q(1-q)(\eta_2 - \eta_1)^2.
\end{equation}

\subsection{Potential Improvement from Mixture Model}
\label{subsec:FI_deterministic}
Let us now demonstrate that exploiting the mixture model has the potential to
improve the ability to estimate SE yield;
this is translated to edge localization in \cref{sec:edge-problem}
after we incorporate source shot noise.
Suppose $m$ ions are incident, with ion $i$ resulting in $X_i$ observed SE counts, for $i=1,\,2,\,\ldots,\,m$.
We will analyze estimators that do and do not use
prior knowledge of $\eta_1$ and $\eta_2$ in the estimation of $\eta = \E{X}$. 

When $q$, $\eta_1$, and $\eta_2$ are all unknown during estimation,
one would naturally estimate $\eta$ using the sample mean of the observations,
\begin{equation}
   \label{eq:etaConv}
     \etaConv = \frac{1}{m}\sum_{i=1}^m X_i.
\end{equation}
The sample mean is unbiased, and by substitution
of \eqref{eq:mixture-variance-two-component},
its variance is
\begin{equation}
  \label{eq:variance-etaConv}
   \variance{\etaConv}
   = \frac{1}{m}\left[
    (1-q)\eta_1 + q\eta_2 + q(1-q)(\eta_2 - \eta_1)^2
    \right]. 
\end{equation}
The {\CR} bound for estimation of $\eta$ \emph{without knowledge of $\eta_1$ and $\eta_2$},
implicitly exploiting the mixture model if possible,
is derived in Supplementary Note~1.
It is numerically indistinguishable from \eqref{eq:variance-etaConv}.

\begin{figure}
  \vspace{\MatlabVerticalTighten}
        \centerline{\includegraphics[width=\singleColumnGraphWidth]{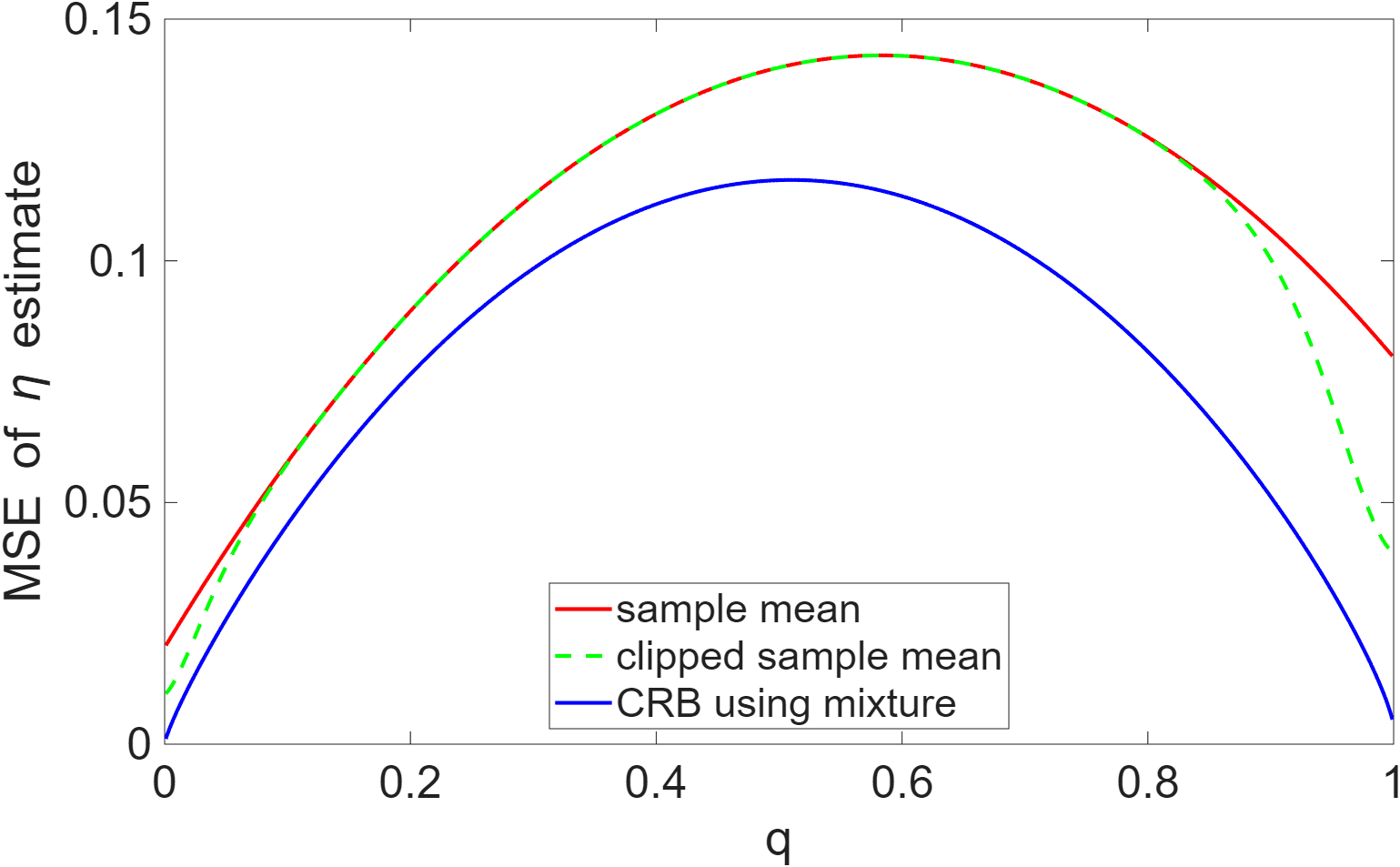}}
  \vspace{\MatlabVerticalTighten}
\caption{Variances in SE yield estimation
    with and without exploiting a mixture model \eqref{eq:X-distribution-two-component}
    and knowledge of $(\eta_1,\eta_2)$
    as a function of mixing parameter $q$.
    Without a distributional model,
    SE yield is estimated with the sample mean,
    which can be clipped to exploit knowledge of $(\eta_1,\eta_2)$.
    With the mixture model, an estimate
    achieving the {\CR} bound would provide an improvement.
    Plotted for $m=100$,
    $\eta_1 = 2$, $\eta_2 = 8$.
    }
\label{fig:mixture-advantage}
\end{figure}

When the values of $\eta_1$, and $\eta_2$ are known,
they provide the constraint of $\eta \in [\eta_1,\,\eta_2]$.
In the absence of a model for the $X_i$ distribution,
it is natural to base an estimate on $\etaConv$,
possibly with some further computation.\footnote{If $X_i$ is Poisson, $\etaConv$ is a sufficient statistic for estimation of $\eta$ from $(X_1,\,X_2,\,\ldots,X_m)$.  Estimation from the sum $\sum_{i=1}^m X_i$ is also justified under the conventional observation model of \cref{subsec:measurement-models}.}
For $m \gtrsim$ 30,
the central limit theorem implies that $\etaConv$ is approximately Gaussian, and
thus clipping $\etaConv$ to the interval $[\eta_1,\,\eta_2$]
gives an approximate constrained maximum likelihood estimate of $\eta$ from $X_1 + X_2 + \cdots + X_m$:
\begin{equation}
    \label{eq:eta_clip}
    \etaClip = \median(\{ \eta_1,\, \etaConv,\, \eta_2 \}).
\end{equation}
Reduction in MSE due to clipping occurs for $q$ near 0 or near 1,
as illustrated in \cref{fig:mixture-advantage}
for $\eta_1 = 2$, $\eta_2 = 8$, and $m = 100$.

In the case of $\eta_1$ and $\eta_2$ being known, we can use the mixture model
to improve the quality of the SE yield estimate.
To derive a {\CR} bound (CRB),
we start by calculating the Fisher information (FI) about $q$ in $X$, with $\eta_1$ and $\eta_2$ given, which we denote
$\Imixture_{X}(q; \eta_1, \eta_2)$.%
\footnote{In our notation, $\mathcal{I}_Z^{\mathrm{model}}(\theta \sMid \phi_1, \phi_2 )$ is the Fisher information in observation $Z$ about parameter $\theta$ when parameters $\phi_1$ and $\phi_2$ are known.  Since the SE yield parameter $\eta$ is applicable to multiple data models, the data model is emphasized with superscript text.}

Using \eqref{eq:mixpoissondistr},
\begin{equation}
\label{eq:dlog_Px_dq}
  \frac{\partial \log \mathrm{P}_X( x \sMid q, \eta_1, \eta_2)}
       {\partial q}
  = \frac{-(\eta_1^x e^{-\eta_1} - \eta_2^x e^{-\eta_2})}
         {(1-q)\eta_1^x e^{-\eta_1} + q\eta_2^x e^{-\eta_2}}.
\end{equation}
Thus, by substitution and simplification,
\begin{align}
    \Imixture_{X}(q \sMid \eta_1, \eta_2)
     &= 
       -\E{ \left( \frac{\partial \log \mathrm{P}_X( X \sMid q, \eta_1, \eta_2)}
                        {\partial q}
            \right)^2 } 
            \nonumber \\
    &= \sum_{x=0}^\infty \frac{1}{x!}
    \frac{\left(
            \eta_1^{x} e^{-\eta_1} - \eta_2^{x} e^{-\eta_2} 
          \right)^2}
         {(1-q) \eta_1^{x} e^{-\eta_1} + q \eta_2^{x} e^{-\eta_2}}.
  \label{eq:I_X_q}
\end{align}
The FI about $\eta$ can be obtained with a well-known property of reparameterization\cite[Ch.~2 (5.11)]{LehmannC:98}.
We divide by
$(\partial \eta/\partial q)^2 = (\eta_2 - \eta_1)^2$
to get
\begin{align}
    \Imixture_{X}(\eta \sMid \eta_1, \eta_2)
    &= \frac{\Imixture_{X}(q \sMid \eta_1, \eta_2)}
            {(\eta_2-\eta_1)^2}.
  \label{eq:I_X_eta}
\end{align}

Since FI is additive,
the reciprocal of
    $m \, \Imixture_{X}(\eta \sMid \eta_1, \eta_2)$
lower bounds the variance of an unbiased estimate for $\eta$
computed from the response to $m$ ions.
This CRB is included in \cref{fig:mixture-advantage}.
The sizeable gaps from the MSEs of $\etaConv$ and $\etaClip$
suggest that using the mixture model---with knowledge of $\eta_1$ and $\eta_2$---provides a potential advantage that 
is not merely due to the knowledge of $\eta_1$ and $\eta_2$.
We seek to realize this advantage.

\subsection{Challenges}
\label{sec:challenges}

The calculations culminating in \cref{fig:mixture-advantage}
isolate the effect of the beam cross section
(randomness in transverse spatial position, see \cref{fig:-pbmmodel}(c))
from the effects of source shot noise and detector noise.
The remainder of the paper studies settings where ion incidences are random
in time (see \cref{fig:-pbmmodel}(b)),
with the aims of
(i) finding theoretical limits of gains analogous to the gap in \cref{fig:mixture-advantage}; and
(ii) developing and demonstrating estimation methods.
The processing of experimental data in \cref{sec:experimental}
will include rudimentary consideration of detector noise.

Recall from \cref{subsec:beam_sample_models} that the number of incident ions in
a fixed dwell time is a random quantity $M$.
As reviewed in
\cref{subsec:measurement-models}---even in the
absence of detector noise---the
effect of source shot noise in classical SE imaging is for the sum of SE counts
$Y = \sum_{i=1}^M X_i$
to be observed rather than the collection
$\{X_1,\,X_2,\,\ldots,\,X_M\}$.
We will extend some previous results on the greater information content in
$\{X_1,\,X_2,\,\ldots,\,X_M\}$
than in $Y$ for the mixture distribution of $X$ in \eqref{eq:X-distribution-two-component}.

There is no observable difference between the lack of an ion and the incidence of an ion that generates zero SEs.
Observing only strictly positive numbers of SEs
leads to zero truncation of the SE count distributions discussed thus far.
Though conceptually simple, this makes the
models substantially more complicated~\cite{Valero2010},
as we will see in \cref{sec:truncated-mix}.

\section{Measurement Models}
\label{sec:TRM}
Peng et al.~\cite{PengMBBG:20} initiated the mitigation of source shot noise through
discrete-time
time-resolved measurement.
Subsequent work~\cite{PengMBG:21}
introduced the abstraction of continuous-time TRM,
which is amenable to more elegant
designs of estimators and
analyses of Fisher information.
These previous works do not include any consideration of beam cross-sectional effects
and thus,
consistent with setting $\beamSigma = 0$ in \cref{subsec:beam_sample_models,subsubsec:sample-char,subsec:se_distribution_one_ion},
treat SE counts as Poisson distributed.

In this section, we review two of the four measurement models from~\cite{PengMBG:21},
omitting the oracle and discrete-time models that we do not build upon in this work;
see~\cite[Sect.~II]{PengMBG:21} for additional details and illustrations.
Then we summarize
basic
analyses for these models,
slightly generalizing beyond
the Poisson distribution for SE counts.

\subsection{Measurement Models with Source Shot Noise}
\label{subsec:measurement-models}
In this section, we adopt the incident beam model in \cref{subsec:beam_sample_models},
but we omit any
consideration of
the beam cross section.
In a fixed dwell time $t$, the number of incident ions is
\begin{equation}
  \label{eq:M-dist}
    M \sim \Poisson(\lambda).
\end{equation}
As illustrated in \cref{fig:-pbmmodel}(b),
ions are incident at times
$(T_1,\,T_2,\,\ldots,\,T_M)$.
Denote the corresponding numbers of generated SEs by
$(X_1,\,X_2,\,\ldots,\,X_M)$.
In~\cite{PengMBBG:20,PengMBG:21,PengKSYG:23,Seidel2022TCI,AgarwalPG:23,Agarwal:2024-PNAS},
each $X_i$ is an independent $\Poisson(\eta)$ random variable.
In general, we will denote
the mean and variance of $X_i$
by $\eta$ and $\sigma_{X}^2$.

Since the case of $X_{i}=0$ produces no detected SEs, the corresponding ion arrival time is not observable in practice.
Omitting such arrivals results in
a thinned, marked Poisson process,
$\{(\Ttilde_1, \Xtilde_1),\,
   (\Ttilde_2, \Xtilde_2),\,
   \ldots,\,
   (\Ttilde_{\Mtilde}, \Xtilde_{\Mtilde})\}$,
where $\Ttilde_i$ is the arrival time of the
$i$th ion that produces a
\emph{positive} number of detected SEs.
The number of ions that produce a positive number of detected SEs is
\begin{equation}
  \label{eq:Mtilde-dist}
    \Mtilde \sim \Poisson(\lambda\rho),
\end{equation}
where
\begin{equation}
    \label{eq:rho-def}
    \rho = \iP{X_i > 0}.
\end{equation}
The distribution of $\Xtilde_i$ is the zero-truncated version of the distribution of $X_i$.

This modeling results in the following measurement abstractions without detector noise:
\begin{itemize}
    \item \emph{Conventional}: Observe only
    \begin{align}
    \label{eq:conventional}
        Y=\sum_{i=1}^M X_i.
    \end{align}
    This is an idealization of the standard operation of existing SE imaging instruments.
    
    \item \emph{Time-resolved measurement}:
    Observe
    \begin{align}
    \label{eq:TRM}
        \{\Mtilde,\,
        (\Ttilde_1,\Xtilde_1),\,
        (\Ttilde_2,\Xtilde_2),\,
        \ldots,\,
        (\Ttilde_{\Mtilde},\Xtilde_{\Mtilde})\}.
    \end{align}
\end{itemize}

\subsection{Analyses of Conventional Observations}
\label{sec:conventional-analyses}
Since $Y$ is generated from a sum of independent and identically distributed random variables,
where the number of terms in the sum is a Poisson random variable,
it is an example of a compound Poisson random variable~\cite{Adelson:66};
it can be called a \emph{Poisson-compounded} version of $X$.
In the special case of $X_i$ being Poisson distributed, $Y$ is said to have the Neyman Type A distribution.

It is straightforward to compute the mean
and variance,
\begin{align}
    \E{Y} &= \E{M} \, \E{X_i} = \lambda \eta,
      \label{eq:Y-mean-general} \\
    \variance{Y}
      &= \lambda \sigma_X^2 + \lambda \eta^2.
      \label{eq:Y-variance-general}
\end{align}
Thus,
\begin{equation}
  \label{eq:etaBaseline}
    \etaBaseline = \Frac{Y}{\lambda} 
\end{equation}
is an unbiased estimate of $\eta$ and has
mean-squared error 
\begin{equation}
  \label{eq:MSE-baseline}
    \mathrm{MSE}(\etaBaseline) = \frac{\sigma_X^2 + \eta^2}{\lambda}.
\end{equation}

While second-order statistics
are
sufficient for the results above,
computing the FI about $\eta$ in $Y$
requires the distribution of $X_i$ to be explicit.
When $X_i$ is Poisson($\eta$) distributed,
\begin{equation}\label{eq:NFI-high-lambda}
  \lim_{\lambda \rightarrow \infty}
    \frac{\IPoisson_Y(\eta \sMid \lambda)}{\lambda}
    = \frac{1}{\eta(1+\eta)}
    = \frac{1}{\eta} - \frac{1}{1+\eta},
\end{equation}
as derived in~\cite[App.~B]{PengMBG:21}.
With the substitution of $\sigma_X^2 = \eta$
in \eqref{eq:MSE-baseline},
comparing with \eqref{eq:NFI-high-lambda}
shows that $\etaBaseline$ is efficient
in the high $\lambda$ limit.
It is close to efficient%
\footnote{For example, \cite[Fig.~2]{PengMBG:21} shows that for $\eta = 3$ and $\lambda \geq 10$, the gap from efficiency is less then 2.3\%.
The behavior is similar to \cref{fig:CompoundedMixture_FI_q} in \cref{subsec:analysis-conventional}.}
for practically useful values of $\lambda$,
so there is little possible benefit in attempting to improve upon
$\etaBaseline$ when only the conventional measurement is available.
This is part of the motivation of TRM\@.

\subsection{Analyses of Time-Resolved Measurements}
\label{sec:Poisson-TRM-analyses}
As noted in \cref{sec:conventional-analyses},
FI computations require the distribution of $X_i$ to be explicit.
We now review some results from~\cite{PengMBG:21,AgarwalPG:23} for the Poisson case
while highlighting a useful intermediate result that applies without regard to the $X_i$ distribution.

The normalized FI about $\eta$ in the TRM \eqref{eq:TRM} was first derived in~\cite{PengMBG:21}:
\begin{equation}
  \frac{1}{\lambda} \ITRMPoisson(\eta \sMid \lambda)
  = \frac{1}{\lambda} \IPoisson_{\Mtilde,\Ttilde_1,\ldots,\Ttilde_{\Mtilde},\Xtilde_1,\ldots,\Xtilde_{\Mtilde}}(\eta \sMid \lambda)
  = \frac{1}{\eta} - e^{-\eta} .
  \label{eq:NFI-TRM}
\end{equation}
The ratio between \eqref{eq:NFI-TRM} and \eqref{eq:NFI-high-lambda} is
\begin{equation}
\label{eq:TRM-gain-Poisson}
\gainPoisson = (\eta+1)(1-\eta e^{-\eta}),
\end{equation}
which varies from $1$ when $\eta = 0$ to $\eta+1$ when $\eta$ is large.
It represents the multiplicative information gain from TRM in the Poisson case.
The analogous gain for Poisson mixtures is derived in \cref{sec:truncated-mix}.

An insightful decomposition from~\cite{AgarwalPG:23}
is presented here because we will use it in \cref{sec:truncated-mix}.
Through steps detailed in~\cite{PengMBG:21},
\begin{equation}
  \label{eq:legacy-FI-decomposition}
  \ITRM(\eta \sMid \lambda)
  = 
  \mathcal{I}_{\Mtilde}(\eta \sMid \lambda)
  + \E{\Mtilde \sMid \lambda} \mathcal{I}_{\Xtilde_i}(\eta).
\end{equation}
The first term is the information in $\Mtilde$,
and the second term is the information in one $\Xtilde_i$ scaled by the average number of these counts;
$\Ttilde_i$ does not appear because the times carry no information about the $\Xtilde_i$ distribution
that is not already present in $\Mtilde$.
This decomposition holds without assuming $X_i$ to be Poisson distributed.
Furthermore,
\begin{equation}
\label{eq:I_Mtilde_eta}
  \IPoisson_{\Mtilde}(\eta \sMid \lambda)
  = \lambda \frac{e^{-\eta}}{e^{\eta}-1}
\end{equation}
as derived in~\cite[Sect.~III-B]{PengMBG:21}
for the Poisson case might be more plainly understood in terms of $\rho$ as follows.
Since $\Mtilde \sim \Poisson(\lambda\rho)$,
it has FI
$\mathcal{I}_{\Mtilde}(\lambda\rho) = 1/(\lambda\rho)$.
Since $\lambda$ is known, this implies
\begin{equation}
\label{eq:I_Mtilde_rho}
  \mathcal{I}_{\Mtilde}(\rho \sMid \lambda) = \lambda/\rho.
\end{equation}
For the Poisson case of $\rho = 1 - e^{-\eta}$,
\eqref{eq:I_Mtilde_eta} and \eqref{eq:I_Mtilde_rho}
are equivalent through reparametrization~\cite[Ch.~2 (5.11)]{LehmannC:98}
because $(\partial \eta/\partial \rho)^2 = e^{2\eta}$.

\section{Analyses for Poisson Mixture Distribution}
\label{sec:truncated-mix}
We now find the normalized Fisher information for the
conventional observation \eqref{eq:conventional}
and
time-resolved measurement \eqref{eq:TRM}
when the SE distribution $X$ is the two-component Poisson mixture \eqref{eq:X-distribution-two-component}.
This combines the spatial and temporal aspects of the beam model in \cref{subsec:beam_sample_models},
specialized for the sample in \cref{sec:two-valued}.
Our goals are to understand how the mixture-model gain suggested by \cref{fig:mixture-advantage}
is modified by the inclusion of source shot noise and,
similarly,
how the TRM gain \eqref{eq:TRM-gain-Poisson}
is modified by the inclusion of beam cross-sectional mixing.

\subsection{Analyses of Conventional Observations}
\label{subsec:analysis-conventional}
Consider the conventional observation \eqref{eq:conventional}
when the SE distribution $X$ is the two-component Poisson mixture \eqref{eq:X-distribution-two-component}.
Using the portions of \cref{sec:conventional-analyses} that are for a general $X$ distribution,
the mean and variance become
\begin{align}
    \E{Y}
      &= \lambda ((1-q)\eta_1 + q\eta_2), \label{eq:E[Y]}\\
    \variance{Y}
      &= \lambda (\eta + q(1-q)(\eta_2 - \eta_1)^2) + \lambda \eta^2, \label{eq:var(Y)}
\end{align}
by substitution of
\eqref{eq:mixture-mean-two-component} in \eqref{eq:Y-mean-general} and
\eqref{eq:mixture-variance-two-component} in \eqref{eq:Y-variance-general};
recall that $\eta = (1-q)\eta_1 + q \eta_2$, making the second expression more compact.
The validity of $\etaBaseline$ as an unbiased estimator holds, and
\eqref{eq:MSE-baseline}
specializes to
\begin{equation}
    \label{eq:MSE-baseline-mixture}
    \mathrm{MSE}(\etaBaseline) = \frac{
      \eta + \eta^2 + q(1-q)(\eta_2-\eta_1)^2
    }{\lambda}.
\end{equation}

Now to understand the theoretical limits for estimation of $q$ from $Y$,
we derive the FI about $q$ in $Y$.
Instead of working purely algebraically with the PMF \eqref{eq:mixpoissondistr},
let us derive the PMF of $Y$ by considering the mechanisms in the generation of $Y$.
Given $M=m$ incident ions,
they must be split as $m-i$ striking the sample where the SE yield is $\eta_1$ and $i$ striking where the SE yield is $\eta_2$
for some $i \in \{0,\,1\,\ldots,\,m\}$.
Since the sum of independent Poisson random variables is a Poisson random variable,
the result is a
Poisson-distributed
number of SEs
with mean
\begin{equation}
   \mixMean_{m,i} = (m-i)\eta_1 + i\eta_2,
   \qquad i = 0,\,1\,\ldots,\,m.
\end{equation}
This occurs with binomial probability
\begin{equation}
   \binProb_{m,i} = \binom{m}{i}(1-q)^{m-i}q^i,
   \qquad i = 0,\,1\,\ldots,\,m.
\end{equation}
We thus have
\begin{equation}
    \mathrm{P}_{Y|M}(y \smid m) 
     = \sum_{i=0}^m \binProb_{m,i} \frac{\mixMean_{m,i}^y}{y!} e^{-\mixMean_{m,i}}.
\end{equation}
Using the law of total probability and the Poisson distribution of $M$ \eqref{eq:M-dist}, we obtain
\begin{equation}
\label{eq:PY}
    \mathrm{P}_{Y}(y) 
    = \sum_{m=0}^\infty \frac{\lambda^m}{m!}e^{-\lambda}
       \sum_{i=0}^m \binProb_{m,i} \frac{\mixMean_{m,i}^y}{y!} e^{-\mixMean_{m,i}}.
\end{equation}

We now proceed to find the FI about $q$ in $Y$.
Dependence on $q$ in $\mathrm{P}_Y(y)$ is only in $\binProb_{m,i}$.
Thus, we compute
$\binProbDiff_{m,i} = \Frac{\partial \binProb_{m,i}}{\partial q}$:
\begin{equation}
    \binProbDiff_{m,i} = \binom{m}{i}\left[ i(1-q)^{m-i}q^{i-1} - (m-i)(1-q)^{m-i-1}q^i \right].
\end{equation}
Using this notation, define
\begin{equation}
\label{eq:numer}
    \numer(y)
    = \pdv{\mathrm{P}_{Y}(y)}{q}
    = \sum_{m=0}^\infty \frac{\lambda^m}{m!}e^{-\lambda}
       \sum_{i=0}^m \binProbDiff_{m,i} \frac{\mixMean_{m,i}^y}{y!} e^{-\mixMean_{m,i}}.
\end{equation}
Summing over the range of $Y$ with weighting by $\mathrm{P}_Y$ to evaluate the expectation in
the definition of FI 
yields
\begin{equation}
    \Imixture_{Y}(q \sMid \eta_1,\eta_2,\lambda ) 
    = \sum_{y=0}^\infty \frac{(L(y))^2}{\mathrm{P}_Y(y)}.
    \label{eq:I_Y_q_sum}
\end{equation}
While \eqref{eq:I_Y_q_sum} is not readily comprehensible,
it can be used to numerically evaluate
$\Imixture_{Y}(q \sMid \eta_1,\eta_2,\lambda)$.

\begin{figure}
  \vspace{\MatlabVerticalTighten}
  \centerline{\includegraphics[width=\singleColumnGraphWidth]{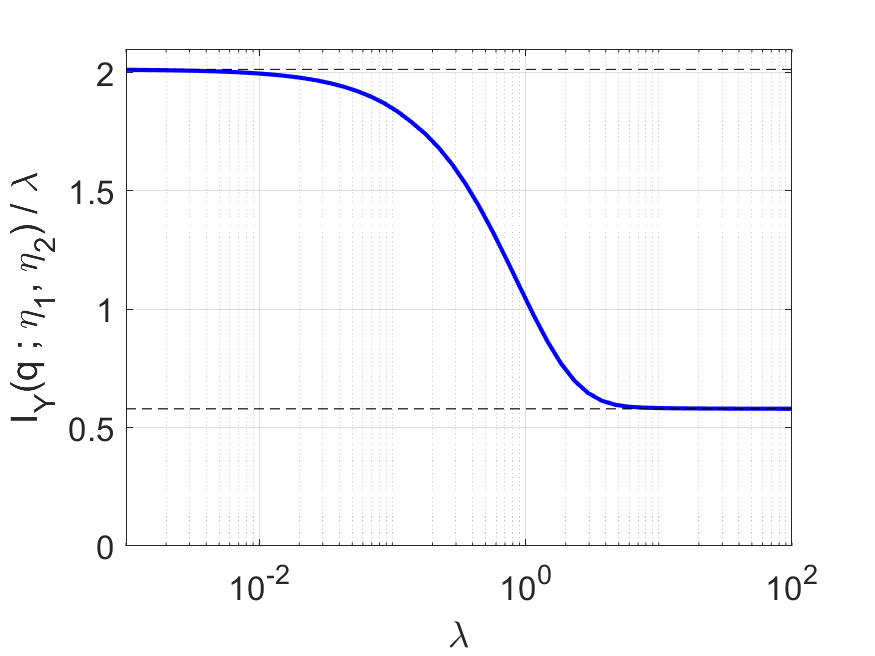}}
  \vspace{\MatlabVerticalTighten}
\caption{Normalized Fisher information $\Imixture_{Y}(q \sMid \eta_1,\eta_2,\lambda) \, / \, \lambda$
  as a function of dose $\lambda$ for $\eta_1=2$, $\eta_2=6$ and $q=0.6$.
  Also shown are the asymptotes
  \eqref{eq:NFI_Y_low_lambda} for low $\lambda$ and
  \eqref{eq:NFI_Y_high_lambda} for high $\lambda$.
  The ratio of the asymptotes is the information gain factor $\gainMixture$ given in \eqref{eq:TRM-gain-mixture}.
  It is the factor by which dose or MSE could be reduced due to the use of time-resolved measurement.}
\label{fig:CompoundedMixture_FI_q}
\end{figure}

At a coarse level,
$\Imixture_{Y}(q \sMid \eta_1, \eta_2, \lambda)$
grows linearly with the dose $\lambda$,
so it is plotted with normalization by $\lambda$ in \cref{fig:CompoundedMixture_FI_q}.
Importantly, the deviation from linearity is for
normalized FI to be a decreasing function of $\lambda$.
Since FI is additive,
this proves that it is advantageous for a measurement at any dose to be replaced by a set of measurements with the same total dose.
This is an argument in favor of time-resolved measurement,
which approaches the behavior of infinitely many measurements, each with infinitesimal $\lambda$~\cite{PengMBG:21}.

The deviations from linearity can be seen from low-
and high-$\lambda$ limits after normalizing by $\lambda$.
The limits
\begin{align}
  \lim_{\lambda \rightarrow 0} &
    \frac{\Imixture_Y(q \sMid \eta_1, \eta_2,\lambda)}{\lambda} 
    = 
    \sum_{y=1}^\infty
    \frac{1}{y!}
    \frac{
    \left(  \eta_1^y e^{-\eta_1}
          - \eta_2^y e^{-\eta_2} \right)^2
    }
    {
           (1-q) \eta_1^y e^{-\eta_1}
            + q  \eta_2^y e^{-\eta_2} 
    } 
  \label{eq:NFI_Y_low_lambda}
\end{align}
and
\begin{align}
    \lim_{\lambda\rightarrow\infty} &
      \frac{\Imixture_Y(q \sMid \eta_1, \eta_2, \lambda)}{\lambda} 
      = \frac{(\eta_2-\eta_1)^2}{\eta + \eta^2 + q(1-q)(\eta_2-\eta_1)^2}
  \label{eq:NFI_Y_high_lambda}
\end{align}
are proven in Supplementary Note~2.
Remarkably, \eqref{eq:NFI_Y_low_lambda} differs from
\eqref{eq:I_X_q} only in the series starting at 1 rather than 0\@.
So this normalized FI is less than
$\Imixture_X(q \sMid \eta_1, \eta_2)$
by
\begin{equation}
\label{eq:FI-gap-X-TRM}
(e^{-\eta_1}-e^{-\eta_2})^2/[(1-q)e^{-\eta_1}+q e^{-\eta_2}].
\end{equation}
This is, in some sense, the cost of having a non-deterministic beam.
By analogy with results for Poisson-distributed $X$~\cite{PengMBG:21},
we would expect the normalized FI for time-resolved measurement
to match low-$\lambda$ asymptote \eqref{eq:NFI_Y_low_lambda};
this is established in \cref{sec:TRM-mixture-analyses}.

For comparisons to \cref{fig:mixture-advantage} and results in \cref{sec:TRM},
we would like the FI about $\eta$ rather than $q$.
Analogously to \eqref{eq:I_X_eta}, we divide by
$(\partial \eta/\partial q)^2 = (\eta_2 - \eta_1)^2$:
\begin{align}
    \label{eq:FI_nontr}
    \Imixture_Y(\eta \sMid \eta_1,\eta_2,\lambda)
    &=\frac{\Imixture_{Y}(q \sMid \eta_1,\eta_2,\lambda)}
    {(\eta_2-\eta_1)^{2}}. 
\end{align}
Applying the same reparametrization to
\eqref{eq:NFI_Y_high_lambda}
gives
\begin{align*}
    \lim_{\lambda\rightarrow\infty} &
      \frac{\Imixture_Y(\eta \sMid \eta_1, \eta_2, \lambda)}{\lambda} 
      = \frac{1}{\eta + \eta^2 + q(1-q)(\eta_2-\eta_1)^2}.
\end{align*}
The reciprocal of this normalized FI matches the MSE
\eqref{eq:MSE-baseline-mixture}
of the baseline estimator $\etaBaseline$,
which shows again---as in \cref{sec:conventional-analyses}---that $\etaBaseline$ is efficient in the high $\lambda$ limit.
Improved estimation at useful doses depends on time-resolved measurement.

\subsection{Analyses of Time-Resolved Measurements}
\label{sec:TRM-mixture-analyses}
We now turn to analyses of the time-resolved measurement \eqref{eq:TRM}.
As discussed in \cref{subsec:measurement-models},
we only observe
an SE count when it is positive,
resulting in \emph{observed} ion count $\Mtilde$ with distribution given in \eqref{eq:Mtilde-dist}
and zero-truncated SE count distribution $\Xtilde$.
Recall from \cref{sec:Poisson-TRM-analyses} that
given $\Mtilde$ and
$\{\Xtilde_1,\Xtilde_2,\ldots,\Xtilde_{\Mtilde}\}$,
the detection times
$\{\Ttilde_1,\Ttilde_2,\ldots,\Ttilde_{\Mtilde}\}$
contain no information about the $X$ distribution.

\begin{figure*}
  \vspace{\MatlabVerticalTighten}
    \centering
    \begin{tabular}{@{}ccc@{}}
      \includegraphics[width=0.318\linewidth]{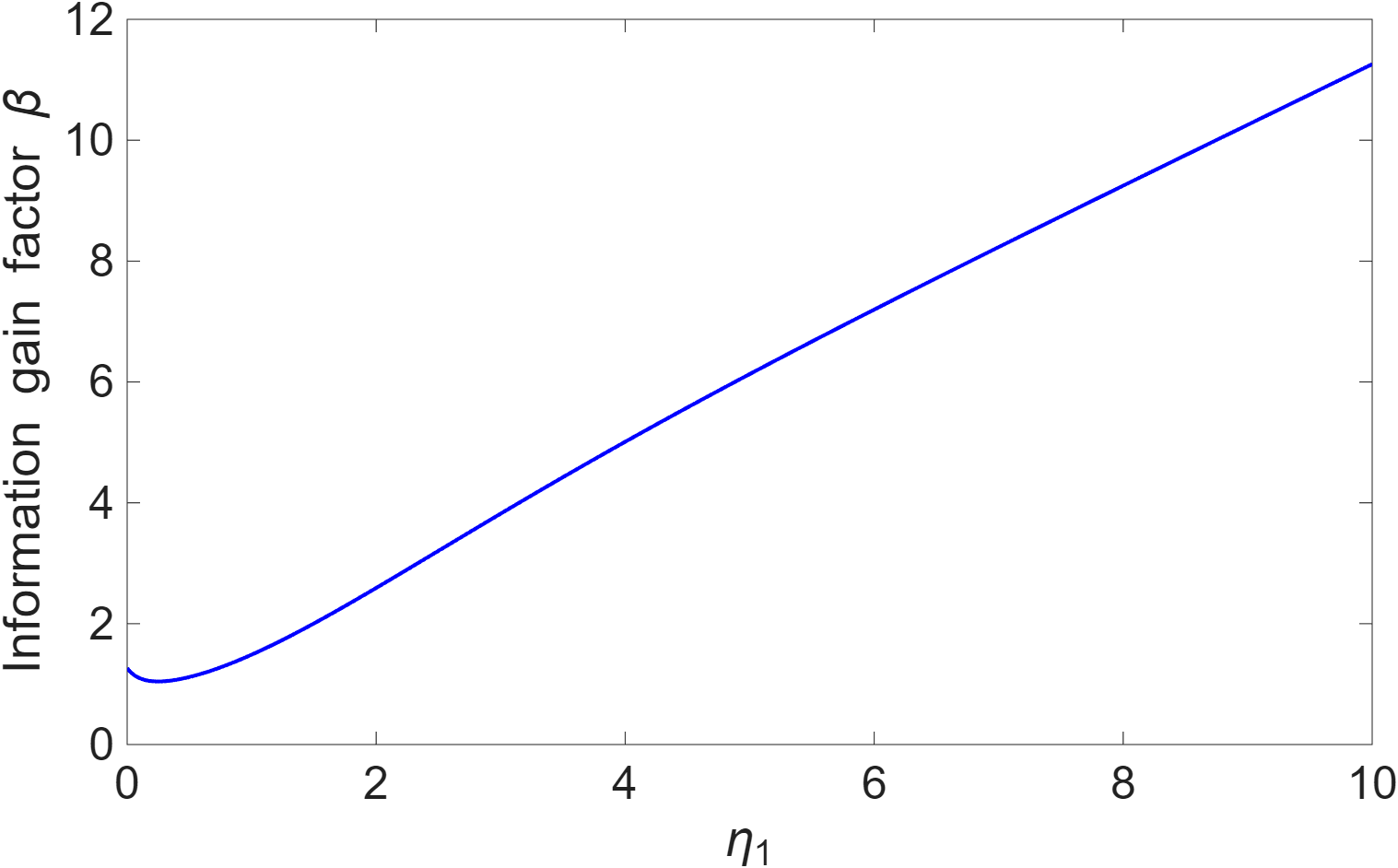} &
      \includegraphics[width=0.318\linewidth]{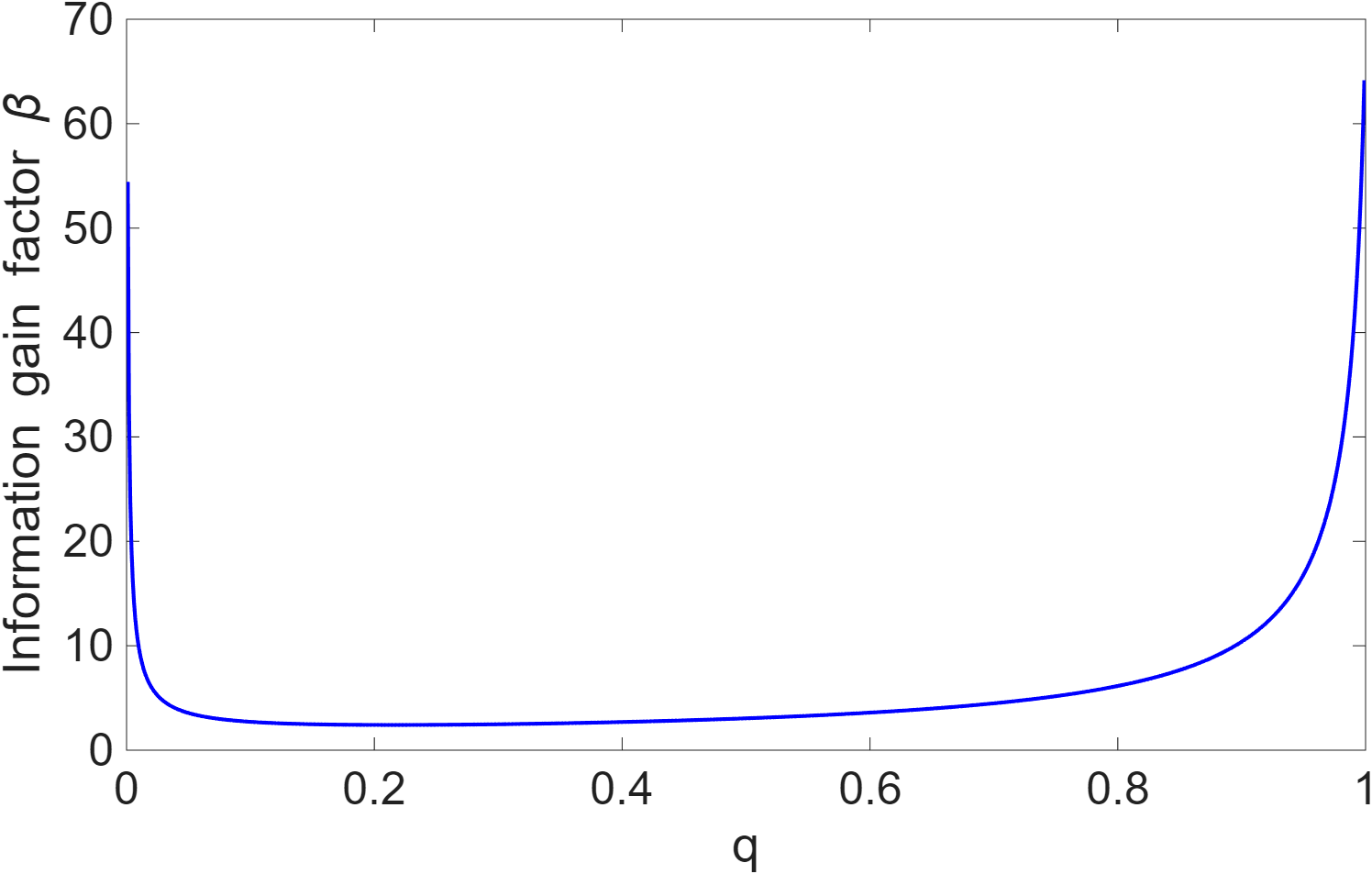} &
      \includegraphics[width=0.318\linewidth]{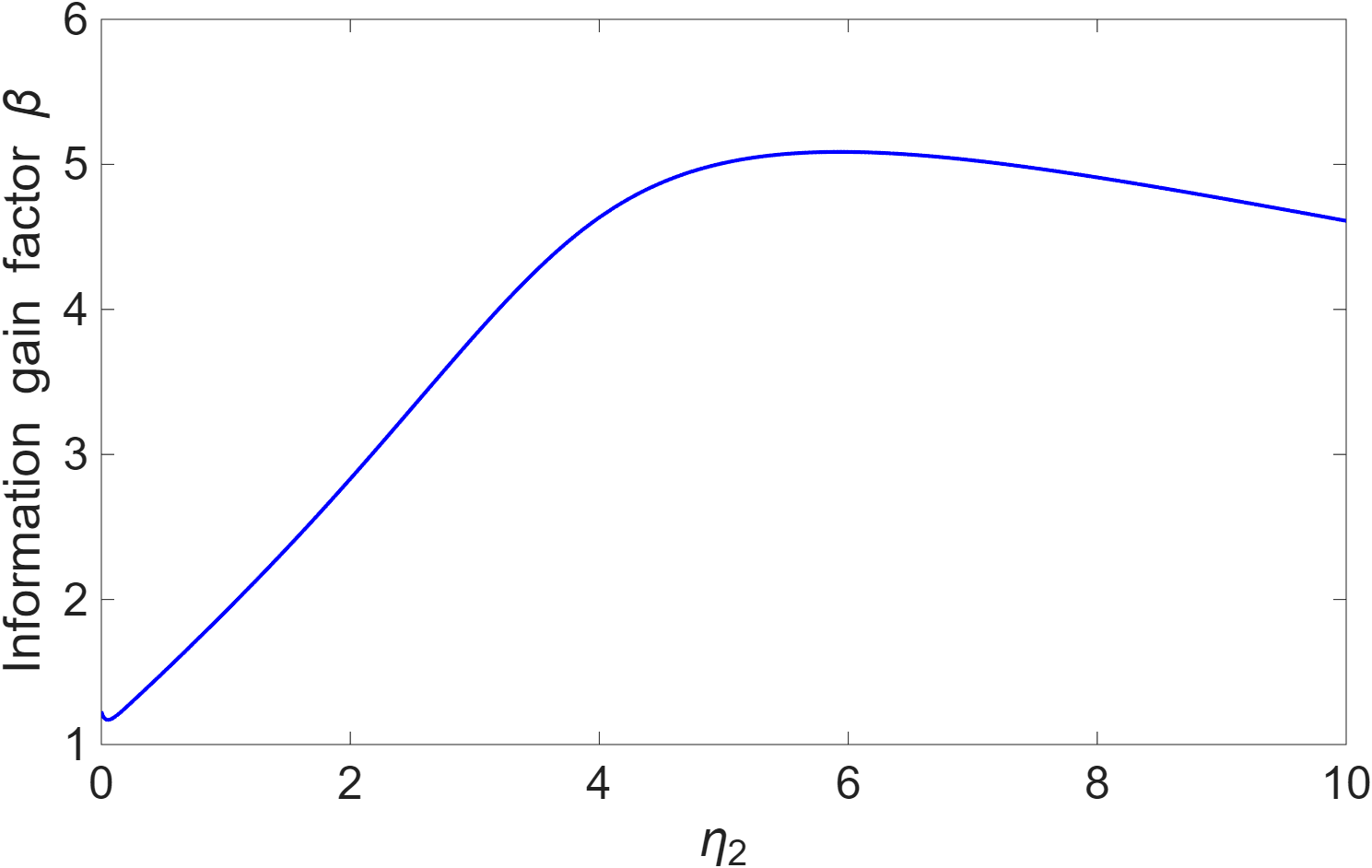} \\
      {\footnotesize (a) $\eta_2=\eta_1+1$, $q=\frac{1}{2}$} &
      {\footnotesize (b) $\eta_1=2$, $\eta_2=8$} &
      {\footnotesize (c) $\eta_1=4$, $q=\frac{1}{2}$}
    \end{tabular}
\caption{
  The information gain factor $\gainMixture$ from time-resolved measurement is the ratio of the FI in TRM to the FI in Poisson-compounded measurement $Y$ at high dose $\lambda$ as given in \eqref{eq:TRM-gain-mixture}.
  With $\eta_1 \approx \eta_2$ as in (a), this gain factor is consistent with
  the $\gainPoisson$ factor for Poisson-distributed $X$ given in \eqref{eq:TRM-gain-Poisson}.}
  \label{fig:ratios-of-asymptotes}
\end{figure*}

Using the Poisson mixture PMF \eqref{eq:mixpoissondistr},
\begin{equation}
    \label{eq:rho-mixture}
    \rho = \iP{X > 0} = 1 - (1-q)e^{-\eta_1} - qe^{-\eta_2}. 
\end{equation}
This $\rho$ and \eqref{eq:mixpoissondistr} give us
\begin{align}
    \mathrm{P}_{\Xtilde}&(\xtilde \sMid q, \eta_1, \eta_2)
    = \mathrm{P}_{X|\{X > 0\}}(\xtilde)
    = \frac{\mathrm{P}_{X}(\xtilde \sMid q, \eta_1, \eta_2)}
           {\rho} \label{eq:ztpm-intermediate} \\
    &= 
    \frac{(1-q)\eta_1^{\xtilde}e^{-\eta_1} + q\eta_2^{\xtilde}e^{-\eta_2}}
         {\xtilde! (1 - (1-q)e^{-\eta_1} - qe^{-\eta_2})},
         \quad \xtilde = 1,\,2,\,\ldots.
         \label{eq:ztpm}
\end{align}
We call this a \emph{zero-truncated Poisson mixture} (ZTPM) distribution.

We can compute the FI about $q$ in TRM with a decomposition analogous to \eqref{eq:legacy-FI-decomposition}:
\begin{align}
  \ITRMmixture(q \sMid & \eta_1, \eta_2, \lambda)
    = \Imixture_{\Mtilde}(q \sMid \eta_1, \eta_2, \lambda)  \nonumber \\
   &+\ \E{\Mtilde \sMid \eta_1, \eta_2, \lambda} \Imixture_{\Xtilde_i}(q \sMid \eta_1, \eta_2).
  \label{eq:q-FI-decomposition}
\end{align}
Two of the three quantities here are very simple.
For the first term,
we have the FI about $\rho$ from \eqref{eq:I_Mtilde_rho}
that can be reparametrized using \eqref{eq:rho-mixture}.
Since
$\Frac{\partial \rho}{\partial q} = e^{-\eta_2}-e^{-\eta_1}$,
we have
\begin{equation}
    \label{eq:FI_MTilde}
    \Imixture_{\Mtilde}(q \sMid \eta_1,\eta_2,\lambda) 
    = \frac{\lambda(e^{-\eta_2}-e^{-\eta_1})^2}{\rho}.
\end{equation}
For the second term, using \eqref{eq:Mtilde-dist},
\begin{equation}
  \label{eq:E_Mtilde}
  \E{\Mtilde \sMid \eta_1, \eta_2, \lambda}
    = \lambda \rho.
\end{equation}
We will compute
\begin{equation}
  \label{eq:I_Xtilde_q_raw}
    \Imixture_{\Xtilde}(q \sMid \eta_1, \eta_2) = \E{
    -\pdv[2]{\log \mathrm{P}_{\Xtilde}(\Xtilde \sMid q, \eta_1, \eta_2)}{q}
    }
\end{equation}
by relating it to earlier computations.
Using \eqref{eq:ztpm-intermediate},
\begin{align}
    -\pdv[2]{\log \mathrm{P}_{\Xtilde}(\xtilde)}{q}
    &= - \pdv[2]{\log \mathrm{P}_{X}(\xtilde)}{q}
       + \pdv[2]{\log \rho}{q}.
\end{align}
Since $\Xtilde$ is the zero-truncated version of $X$,
the expected value of the first term is \eqref{eq:I_X_q} with the first term omitted:
\begin{equation}
    \E{ - \pdv[2]{\log \mathrm{P}_{X}(\Xtilde)}{q} }
    =
    \sum_{\xtilde=1}^\infty \frac{1}{\xtilde!}
    \frac{\left(
            \eta_1^{\xtilde} e^{-\eta_1} - \eta_2^{\xtilde} e^{-\eta_2} 
          \right)^2}
         {(1-q) \eta_1^{\xtilde} e^{-\eta_1} + q \eta_2^{\xtilde} e^{-\eta_2}}.
\end{equation}
Using \eqref{eq:rho-mixture},
\begin{equation}
  \label{eq:rho-second-derivative}
    \pdv[2]{\log \rho}{q}
    = \frac{-(e^{-\eta_2}-e^{-\eta_1})^2}
           {\rho^2},
\end{equation}
which is a deterministic quantity.
Combining \eqref{eq:I_Xtilde_q_raw}--\eqref{eq:rho-second-derivative},
\begin{align}
    &\Imixture_{\Xtilde}(q \sMid \eta_1, \eta_2) \nonumber \\
    &= 
    \left(
    \sum_{\xtilde=1}^\infty \frac{1}{\xtilde!}
    \frac{\left(
            \eta_1^{\xtilde} e^{-\eta_1} - \eta_2^{\xtilde} e^{-\eta_2} 
          \right)^2}
         {(1-q) \eta_1^{\xtilde} e^{-\eta_1} + q \eta_2^{\xtilde} e^{-\eta_2}}
    \right)
    - \frac{(e^{-\eta_2}-e^{-\eta_1})^2}{\rho^2}.
  \label{eq:I_Xtilde_q_final}
\end{align}
Finally, substituting \eqref{eq:FI_MTilde}, \eqref{eq:E_Mtilde}, and \eqref{eq:I_Xtilde_q_final}
into \eqref{eq:q-FI-decomposition} gives
\begin{equation}
  \label{eq:I_TRM}
  \ITRMmixture(q \sMid \eta_1, \eta_2, \lambda) 
  = \lambda
    \sum_{\xtilde=1}^\infty \frac{1}{\xtilde!}
    \frac{\left(
            \eta_1^{\xtilde} e^{-\eta_1} - \eta_2^{\xtilde} e^{-\eta_2} 
          \right)^2}
         {(1-q) \eta_1^{\xtilde} e^{-\eta_1} + q \eta_2^{\xtilde} e^{-\eta_2}}.
\end{equation}

As anticipated,
normalized FI
$\ITRMmixture(q \sMid \eta_1, \eta_2, \lambda) / \lambda$
is identical to the low-$\lambda$ asymptote \eqref{eq:NFI_Y_low_lambda}.
This again theoretically supports the information gain from using TRM\@.
In the case of Poisson-distributed SE counts,
the multiplicative information gain has a simple and interpretable form given
by $\gainPoisson$
in 
\eqref{eq:TRM-gain-Poisson}.
We can similarly define a multiplicative information gain
for the mixture SE distribution as the ratio between \eqref{eq:NFI_Y_low_lambda} and \eqref{eq:NFI_Y_high_lambda}:%
\footnote{While $\gainPoisson$ is a ratio of FIs about $\eta$, here we use FIs about $q$.  Under our assumption that $\eta_1$ and $\eta_2$ are known, these differ from FIs about $\eta$ by a multiplicative factor that cancels in the ratio.}
\begin{align}
    \gainMixture
    &= \frac{
    \sum_{\xtilde=1}^\infty \frac{1}{\xtilde!}
    \frac{\left(
            \eta_1^{\xtilde} e^{-\eta_1} - \eta_2^{\xtilde} e^{-\eta_2} 
          \right)^2}
         {(1-q) \eta_1^{\xtilde} e^{-\eta_1} + q \eta_2^{\xtilde} e^{-\eta_2}}
    }
    {
     \left(\frac{(\eta_2-\eta_1)^2}{\eta + \eta^2 + q(1-q)(\eta_2-\eta_1)^2}\right)
    } \nonumber \\
    &=
    {
     \left(\!
     \frac
     {\eta + \eta^2}
     {(\eta_2-\eta_1)^2}
     + q(1-q)
     \!\right)
    } \nonumber \\
    & \qquad
    {
    \sum_{\xtilde=1}^\infty \frac{1}{\xtilde!}
    \frac{\left(
            \eta_1^{\xtilde} e^{-\eta_1} - \eta_2^{\xtilde} e^{-\eta_2} 
          \right)^2}
         {(1-q) \eta_1^{\xtilde} e^{-\eta_1} + q \eta_2^{\xtilde} e^{-\eta_2}}
    }.
  \label{eq:TRM-gain-mixture}
\end{align}
While it is obvious that $\gainPoisson$ in \eqref{eq:TRM-gain-Poisson}
is approximately linear in $\eta$,
\eqref{eq:TRM-gain-mixture} is more difficult to understand.
We examine it through three plots in \cref{fig:ratios-of-asymptotes}.
\Cref{fig:ratios-of-asymptotes}(a) shows $\gainMixture$ when varying $\eta_1$ with $\eta_2 = \eta_1 + 1$ and $q = \frac{1}{2}$.
The increase of $\gainMixture$ with increasing $\eta_1$ is quite similar to the dependence of $\gainPoisson$ on $\eta$.
In fact, $\gainMixture$ approaches $\gainPoisson$ as $\eta_1$ and $\eta_2$ approach $\eta$.
Cases where the mixture is farther from approximately Poisson are less intuitive.
\Cref{fig:ratios-of-asymptotes}(b) varies $q$ with $\eta_1 = 2$ and $\eta_2 = 8$.
Here the information gain from TRM
$\gainMixture$
increases sharply for
$q \rightarrow 0$
and 
$q \rightarrow 1$.
Finally, \cref{fig:ratios-of-asymptotes}(c) varies $\eta_2$ with $\eta_1 = 4$ and $q = \frac{1}{2}$.

\begin{figure}
  \vspace{\MatlabVerticalTighten}
        \centerline{\includegraphics[width=\singleColumnGraphWidth]{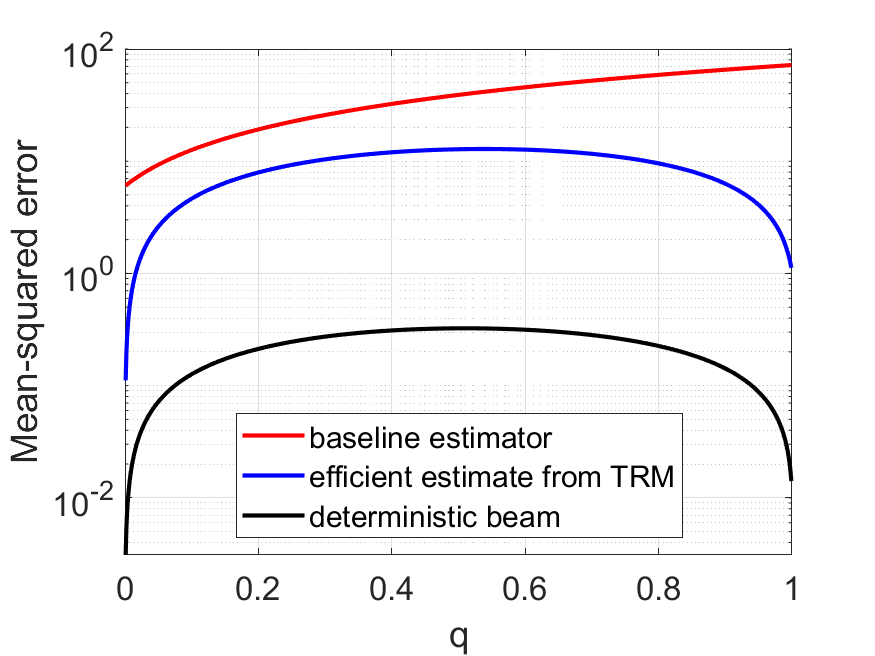}}
  \vspace{\MatlabVerticalTighten}
\caption{MSEs in SE yield estimation
    when SE counts follow the mixture model \eqref{eq:X-distribution-two-component},
    as a function of mixing parameter $q$,
    for $\eta_1 = 2$, $\eta_2 = 8$, and $\lambda = 1$.
    The baseline estimator $\etaBaseline$ \eqref{eq:etaBaseline}
    does not use TRM and hence also cannot exploit the mixture model.
    Performance of an efficient estimate from TRM is described by \eqref{eq:I_TRM} (with appropriate scaling).
    For additional context, performance described by \eqref{eq:I_X_q}
    represents efficient estimation and a deterministic incident beam with $m=1$ ion.
    TRM enables substantial improvements
    as compared to the baseline estimator;
    a deterministic beam would enable further improvements,
    as described by the FI gap \eqref{eq:FI-gap-X-TRM}.
    }
\label{fig:compounded-mixture-advantage}
\end{figure}

In this paper,
a Poisson distribution for $X$ often symbolizes na\"ivety because
(a) it arises from the convolutional model that we argue to be inaccurate; and
(b) it sometimes justifies the uncritical use of sample mean estimators.
The variance reduction
depicted in \cref{fig:mixture-advantage}
is for exploiting or not exploiting 
the mixture model when the mixture model does
hold in describing the data.
We can extend that result from the idealization
of a deterministic beam in \cref{subsec:FI_deterministic}
to the more realistic case with source shot noise by
comparing the MSE of the baseline estimator $\etaBaseline$
to the MSE of an efficient estimator,
i.e., the reciprocal of
$\ITRMmixture(\eta \sMid \eta_1,\eta_2,\lambda)$.
The red and blue curves in
\cref{fig:compounded-mixture-advantage}
provide this comparison for
mean dose
$\lambda = 1$.
The black curve is the MSE lower bound implied by \eqref{eq:I_X_eta}
for deterministically having $m=1$ incident ion.
In the $q \rightarrow 0$ and $q \rightarrow 1$ limits,
the gap between the red and blue curves exactly matches
$\gainPoisson$ from \eqref{eq:TRM-gain-Poisson}.
The gap between the blue and black curves is
explained precisely by
the FI gap
\eqref{eq:FI-gap-X-TRM}.

\section{Edge Localization}
\label{sec:edge-problem}

Recall that \eqref{eq:mixing-q} in \cref{sec:two-valued} connects a mixture parameter $q$ to an edge location $\gamma$ in a two-valued sample as shown in \cref{fig:twoValuedSample}(a).
This type of two-valued sample is inspired by semiconductor inspection problems, where an edge could be between a metal nanostructure and a silicon substrate.
Having established several results in \cref{sec:truncated-mix} on the estimation of $q$ in the presence of source and target shot noise,
we are now ready to study the estimation of $\gamma$---theoretically, in Monte Carlo simulations, and through processing of experimental data from an HIM\@.
The precision of such \emph{edge localization} has implications for the feature size tolerances of nanofabricated electronic devices, with consequent effects on device performance and reliability~\cite{IRDS2023}.
The localization here is not limited to the grid scan locations and
provides sub-pixel accuracy.

In \cref{subsec:FI_gamma}, we use \eqref{eq:I_TRM} to find the Fisher information about the edge location $\gamma$ in data from a collection of raster scan locations.
We also explore optimizing the beam width $\beamSigma$ to minimize the localization error, in worst case among all potential edge locations.
In \cref{sec:edge-estimates},
we introduce four edge location estimates.
The maximum likelihood estimator for $\gamma$ from time-resolved measurement data under the mixture distribution \eqref{eq:X-distribution-two-component} for SE counts $X$
is our central contribution for edge localization,
and we wish to compare it to several alternatives.
Current practice is represented by linear interpolation of the baseline estimate of SE yield at each scan location.
This is slightly improved by clipping as in \cref{subsec:FI_deterministic}; since
$\eta_1$ and $\eta_2$ known, SE yield estimates are clipped to $[\eta_1,\eta_2]$,
eliminating the cases in which $\etaBaseline$ is slightly less than $\eta_1$ or slightly greater than $\eta_2$\@.
We also introduce an MLE formulation for TRM data under the Poisson distribution of the data,
i.e., treating the effect of the beam cross-section as a convolution;
this is \emph{mismatched} with respect to the more accurate mixture model.
Comparisons based on Monte Carlo simulations appear in \cref{sec:simresults}.
Finally, \cref{sec:experimental} uses experimental data acquired at an edge between gold and silicon on a helium ion microscope to show the same trends in estimator performance as the Monte Carlo simulations.

\subsection{Fisher Information about the Edge Location in TRM}
\label{subsec:FI_gamma}
Using \eqref{eq:mixing-q} for the relationship between mixing parameter $q$ and edge location $\gamma$ when the scan location is $g_1$,
\begin{equation}
   \pdv{q}{\gamma}
     = - \frac{ \Phi'( (\gamma-g_1)/\beamSigma ) }{ \beamSigma }
     = - \frac{e^{-\frac{1}{2}((\gamma-g_1)/\beamSigma)^2}}{\sqrt{2\pi} \, \beamSigma}.
\end{equation}
Then the reparameterization
of
 \eqref{eq:I_TRM}
 gives Fisher information
\begin{align}
  \left( \lambda
    \sum_{\xtilde=1}^\infty \frac{1}{\xtilde!}
    \frac{\left(
            \eta_1^{\xtilde} e^{-\eta_1} - \eta_2^{\xtilde} e^{-\eta_2} 
          \right)^2}
         {(1-q) \eta_1^{\xtilde} e^{-\eta_1} + q \eta_2^{\xtilde} e^{-\eta_2}}
  \right)
         \frac{e^{{\Frac{-(\gamma-g_1)^2}{\beamSigma^2}}}}{{2\pi}\beamSigma^2}
\end{align}
about $\gamma$.
We consider scan locations on one horizontal line at positions $\{0,\,1,\ldots,\,\ell-1\}$.
Using additivity of FI, we have FI
\begin{align}
    &\Iscanmixture(\gamma \sMid \eta_1,\eta_2,\lambda,\beamSigma) \nonumber \\
    &=
  \sum_{k=0}^{\ell-1}
  \left( \lambda
    \sum_{\xtilde=1}^\infty \frac{1}{\xtilde!}
    \frac{\left(
            \eta_1^{\xtilde} e^{-\eta_1} - \eta_2^{\xtilde} e^{-\eta_2} 
          \right)^2}
         {(1-q_k) \eta_1^{\xtilde} e^{-\eta_1} + q_k \eta_2^{\xtilde} e^{-\eta_2}}
  \right)
         \frac{e^{{\Frac{-(\gamma-k)^2}{\beamSigma^2}}}}{{2\pi}\beamSigma^2}
    \label{eq:FI_gamma}
\end{align}
for the full scan, where $q_k = 1 - \Phi((\gamma-k)/\beamSigma)$ from \eqref{eq:mixing-q}.

\begin{figure}
  \vspace{\MatlabVerticalTighten}
 \centering
  \begin{tabular}{c}  
    \includegraphics[width=\singleColumnGraphWidth]{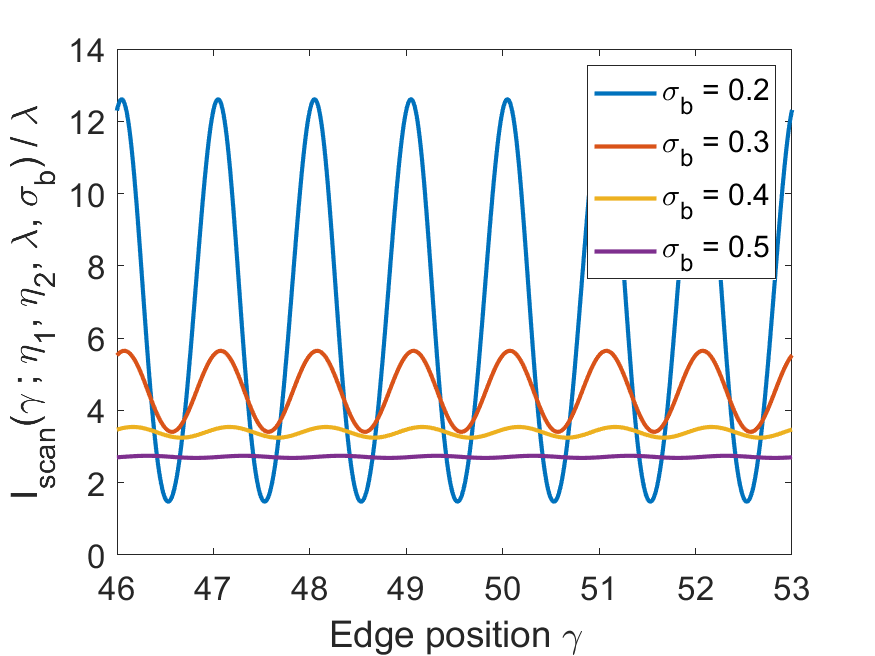} \\
    {\footnotesize (a)
  $\Iscanmixture(\gamma \sMid \eta_1,\eta_2,\lambda, \beamSigma) / \lambda$
  as a function of $\gamma$
    } \\
    \includegraphics[width=\singleColumnGraphWidth]{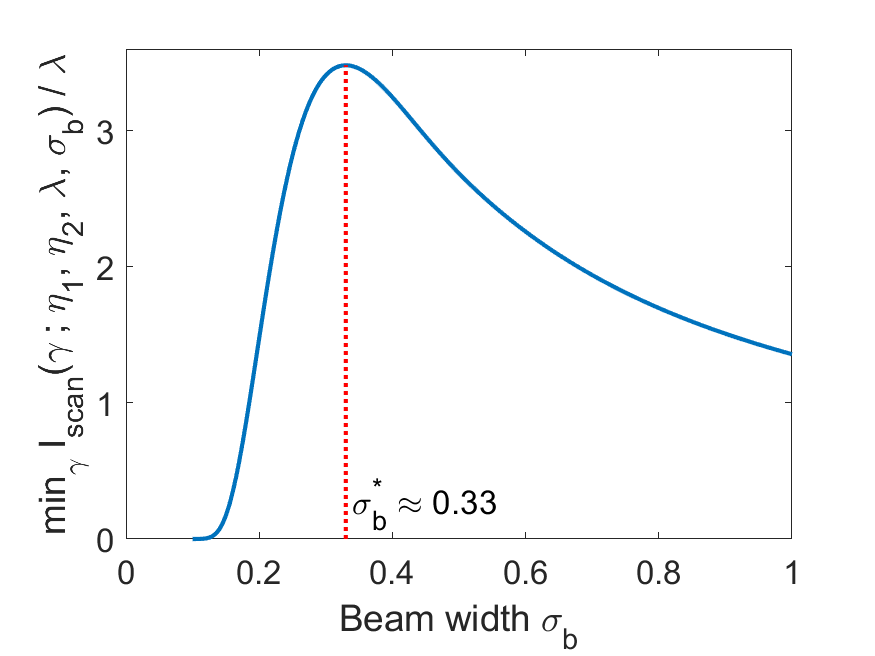} \\
    {\footnotesize (b)
  $\min_\gamma \Iscanmixture(\gamma \sMid \eta_1,\eta_2,\lambda, \beamSigma) / \lambda$
  as a function of $\beamSigma$
    } 
  \end{tabular}
  \caption{Study of Fisher information about the edge location $\gamma$ in a full scan at $\{0,\,1,\,\ldots,\,99\}$.
  (a) Dependence of dose-normalized FI
  on $\gamma$
  (plotted for $\eta_1=1$, $\eta_2=10$, and four values of beam width $\beamSigma$).
  FI is highest when the edge is near a scan grid point and lowest when the edge is near the midpoint between scan grid points.
  (b) Dependence of the minimum over $\gamma$ of the dose-normalized FI on $\beamSigma$.
  The maximin optimal value of the beam width is $\beamSigma^\star=0.33$.
    }
\label{fig:fi_gamma_variation}
\end{figure}

Consistent with many results in earlier sections,
$\Iscanmixture(\gamma \sMid \eta_1,\eta_2,\lambda,\beamSigma)$
is proportional to the dose $\lambda$;
thus, we further study this FI normalized by $\lambda$.
The dependence on beam width $\beamSigma$ and edge location $\gamma$ is more intricate. 
In \cref{fig:fi_gamma_variation}(a), we plot
$\Iscanmixture(\gamma \sMid \eta_1,\eta_2,\lambda,\beamSigma)/\lambda$
for $\gamma$ varied continuously between 46 and 53
for four values of $\beamSigma$. 
The normalized FI exhibits oscillations
depending on proximity to the scan grid.
The oscillations are larger when the beam is narrow
(e.g., $\beamSigma \in \{0.2,\,0.3\}$)
because an edge location $\gamma$ far from the scan grid
(e.g., at 49.5)
will result in a very small fraction of incident ions striking near the edge,
while the opposite will happen when $\gamma$ is near the scan grid.
Increasing the beam width
(e.g., $\beamSigma \in \{0.4,0.5\}$)
dampens the oscillations but also reduces the normalized FI\@.

Since the edge location is not known \emph{a priori},
it is desirable to maximize the informativeness of the measurements for the worst-case edge location.
Hence, we define an optimal beam width $\beamSigma^{\star}$
through a maximin formulation:
\begin{align}
    \beamSigma^{\star} =
    \argmax_{\beamSigma} \, \min_{\gamma \in \Gamma} \Iscanmixture(\gamma \sMid \eta_1,\eta_2,\lambda,\beamSigma)/\lambda,
\end{align}
where $\Gamma$ is the set of possible edge locations.
For $(\eta_1,\eta_2)=(1,10)$,
\cref{fig:fi_gamma_variation}(b) shows the minimum normalized FI for a range of beam widths, and it shows the maximum occurring at
$\beamSigma^\star = 0.33$. 
This is not a universal constant. The dependence of $\beamSigma^\star$ on $(\eta_1,\eta_2)$ is mild,
as illustrated in Supplementary Note~3.

\subsection{Edge Location Estimates}
\label{sec:edge-estimates}
\emph{Interpolation:}
The current practice of SE imaging is based on computing the baseline estimate \eqref{eq:etaBaseline} at each pixel prior to any subsequent processing.
If the knowledge of $\eta_1$ and $\eta_2$ is available a priori, one can clip this $\etaBaseline$ to lie in $[\eta_1,\eta_2]$.
In the absence of an explicit model for the effect of the beam cross section,
one could declare an edge location to be where the estimated SE yield crosses a threshold value. 
For example, the edge location could be estimated at the location where the linearly interpolated estimated SE yield equals $(\eta_1+\eta_2)/2$\@.
We refer to the edge estimator based on level crossings of $\etaBaseline$ as the \emph{Interpolation} estimator,
while the estimator that uses the clipped version of $\etaBaseline$ as the \emph{Interpolation-clip} estimator.
In our specific case of left-to-right scanning from lower SE yield to higher SE yield,
as illustrated in \cref{fig:twoValuedSample}(b),
the interpolated SE yield estimate function may cross the threshold multiple times. 
When this happens, we use the sample mean of all the upward crossings.

\emph{Maximum likelihood estimate:}
Formulating the MLE of $\gamma$ from a full scan of TRM data,
$\{ (\mtilde_k,\{\xtilde_{i,k}\}_{i=1}^{\mtilde_k}) \}_{k=0}^{\ell-1}$,
where $k$ indexes the scan location,
is a matter of collecting the likelihood expressions from earlier sections.
For any candidate edge location $\gamma$ and each horizontal raster scan location  $k$,
\eqref{eq:mixing-q}
provides the corresponding mixing parameter  $q_k$.
The likelihood of observing
$\mtilde_k$
is given by the Poisson distribution with parameter
 $\lambda \rho_k$,
where
$\rho_k$
is given by \eqref{eq:rho-mixture}; and
the likelihood for each $\xtilde_{i,k}$ is given by \eqref{eq:ztpm}.
Combining these over all $\ell$ raster scan positions gives the likelihood of the full data set,
\begin{equation}
  \label{eq:LScan}
    \Lscanmixture(\gamma) = \prod_{k=0}^{\ell-1} \Poisson(\mtilde_k \sMid \lambda \rho_k)
                     \prod_{i=1}^{\mtilde_k} 
    \mathrm{P}_{\Xtilde}(\xtilde_{i,k} \sMid q_k, \eta_1, \eta_2),
\end{equation}
where $\rho_k$ and $q_k$ are functions of $\gamma$, $k$, $\eta_1$, $\eta_2$, and $\beamSigma$.
Maximizing this scalar function of $\gamma$ gives the MLE:
\begin{equation}
  \label{eq:gammaMLE}
   \gammaMLE = \argmax_\gamma \Lscanmixture(\gamma).
\end{equation}

\emph{Mismatched MLE:}
To isolate the effect of the mixture model
 from the effect of TRM, we also consider an estimator that maximizes the likelihood for Poisson-distributed data;
i.e., the estimator treats the effect of the beam cross-section as a convolution.
As above,
for any candidate edge location $\gamma$ and each horizontal raster scan location $k$,
we have a corresponding parameter $q_k$ from \eqref{eq:mixing-q}.
This implies a mean $\mu_k$ given by 
\eqref{eq:mixture-mean-two-component},
and each SE count distribution $X_{i,k}$ is assumed to be $\Poisson(\mu_k)$.
Differing from above, the thinning parameter in the distribution of $\Mtilde$ is $\rho_k' = 1-e^{-\mu_k}$.
Also differing from above, each $\Xtilde_{i,k}$ has a zero-truncated Poisson distribution
with parameter $\mu_k$, $\ZTPoisson(\mu_k)$.
Combining the likelihoods over the scan positions gives
\begin{align*}
  \LscanPoisson&(\gamma) \\
                   &= \prod_{k=0}^{\ell-1} \Poisson(\mtilde_k \sMid \lambda \rho_k')
                     \prod_{i=1}^{\mtilde_k} 
    \ZTPoisson(\xtilde_{i,k} \sMid \mu_k ),
\end{align*}
where $\rho_k'$ and $\mu_k$ are functions of $\gamma$, $k$, $\eta_1$, $\eta_2$, and $\beamSigma$.
Maximizing this function gives the mismatched MLE (MMLE):
\begin{equation}
  \label{eq:gammaMMLE}
   \gammaMMLE = \argmax_\gamma \LscanPoisson(\gamma).
\end{equation}

\newlength{\tableFigureWidth}
\setlength{\tableFigureWidth}{0.649\linewidth}
\newlength{\tightenRow}
\setlength{\tightenRow}{-1.2mm}
\newlength{\rowSkip}
\setlength{\rowSkip}{3mm}

\begin{figure*}[t!]
 \centering
  \begin{tabular}{@{}ccc@{}}
    {\small Bias} & {\small Standard deviation} & {\small Root mean-squared error} \\
    \includegraphics[width=\tableFigureWidth]{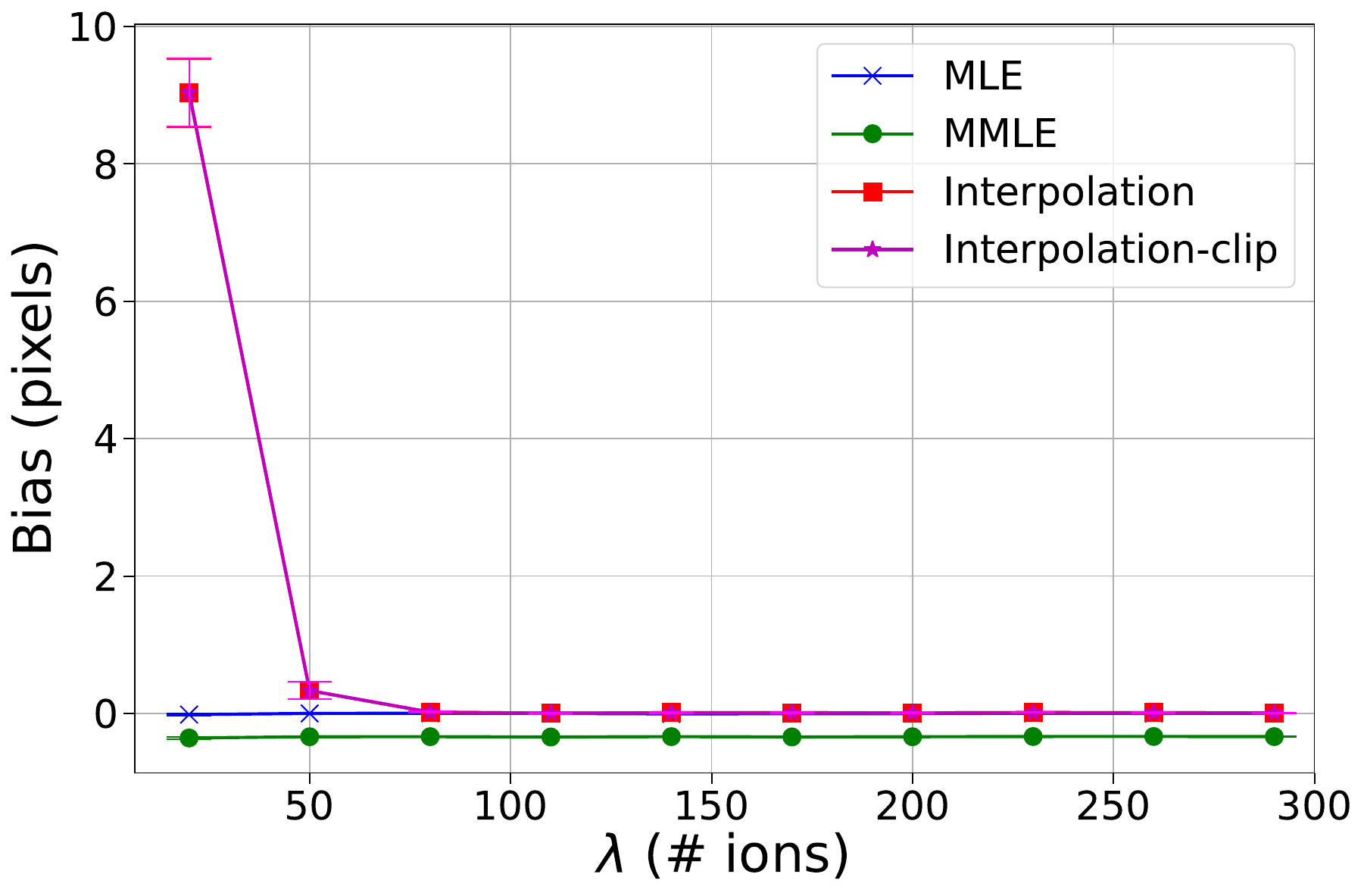} &
    \includegraphics[width=\tableFigureWidth]{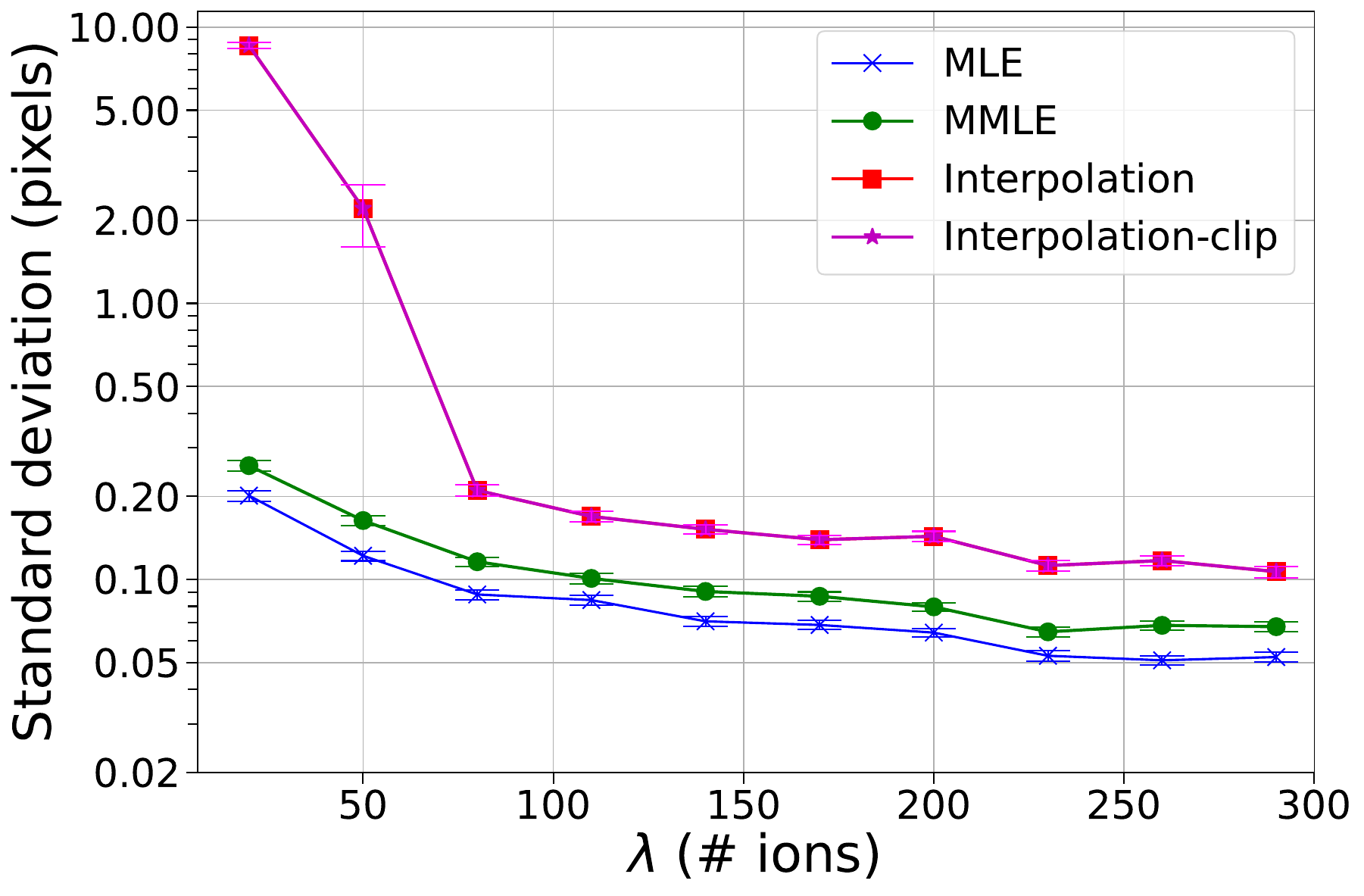} &
    \includegraphics[width=\tableFigureWidth]{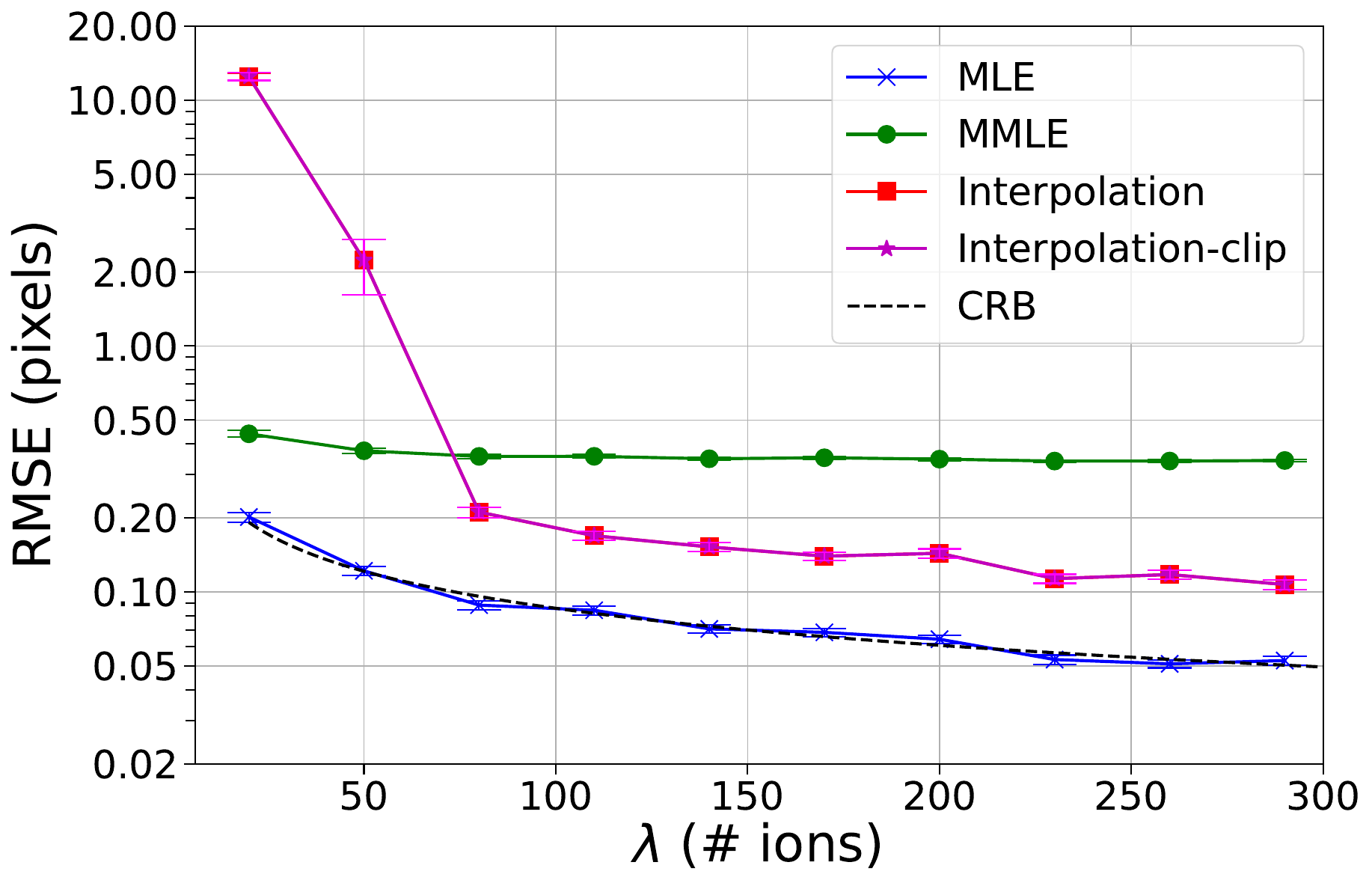} \\[\tightenRow]
    \multicolumn{3}{l}{\footnotesize (a) Varying dose $\lambda$ from 20 to 290 with
    edge location $\gamma = 50.2$,
    beam width $\beamSigma=1$,
    and sample SE yields $\eta_1=1$ and $\eta_2=10$.}
    \\[\rowSkip]
    
    \includegraphics[width=\tableFigureWidth]{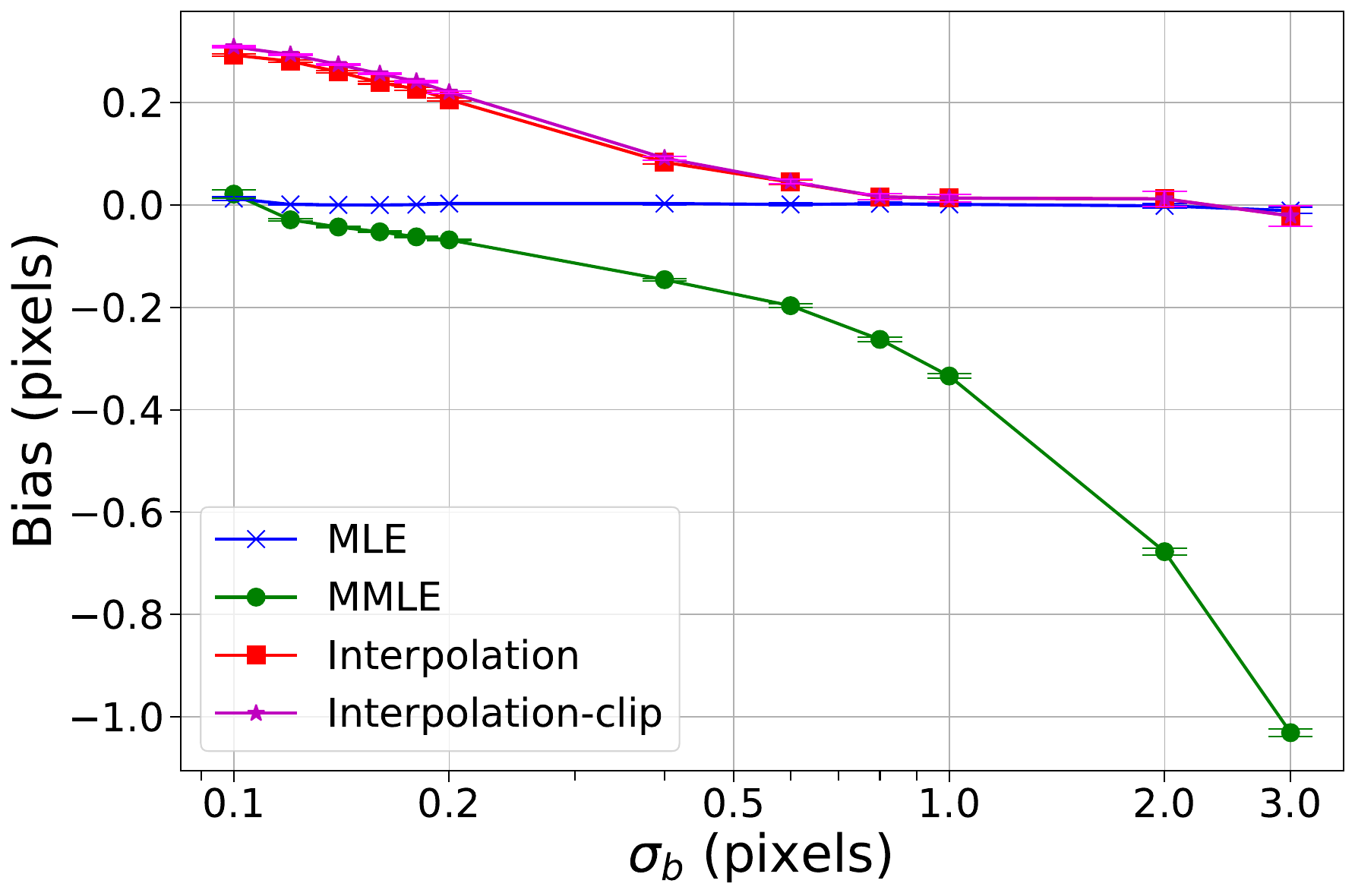} &
    \includegraphics[width=\tableFigureWidth]{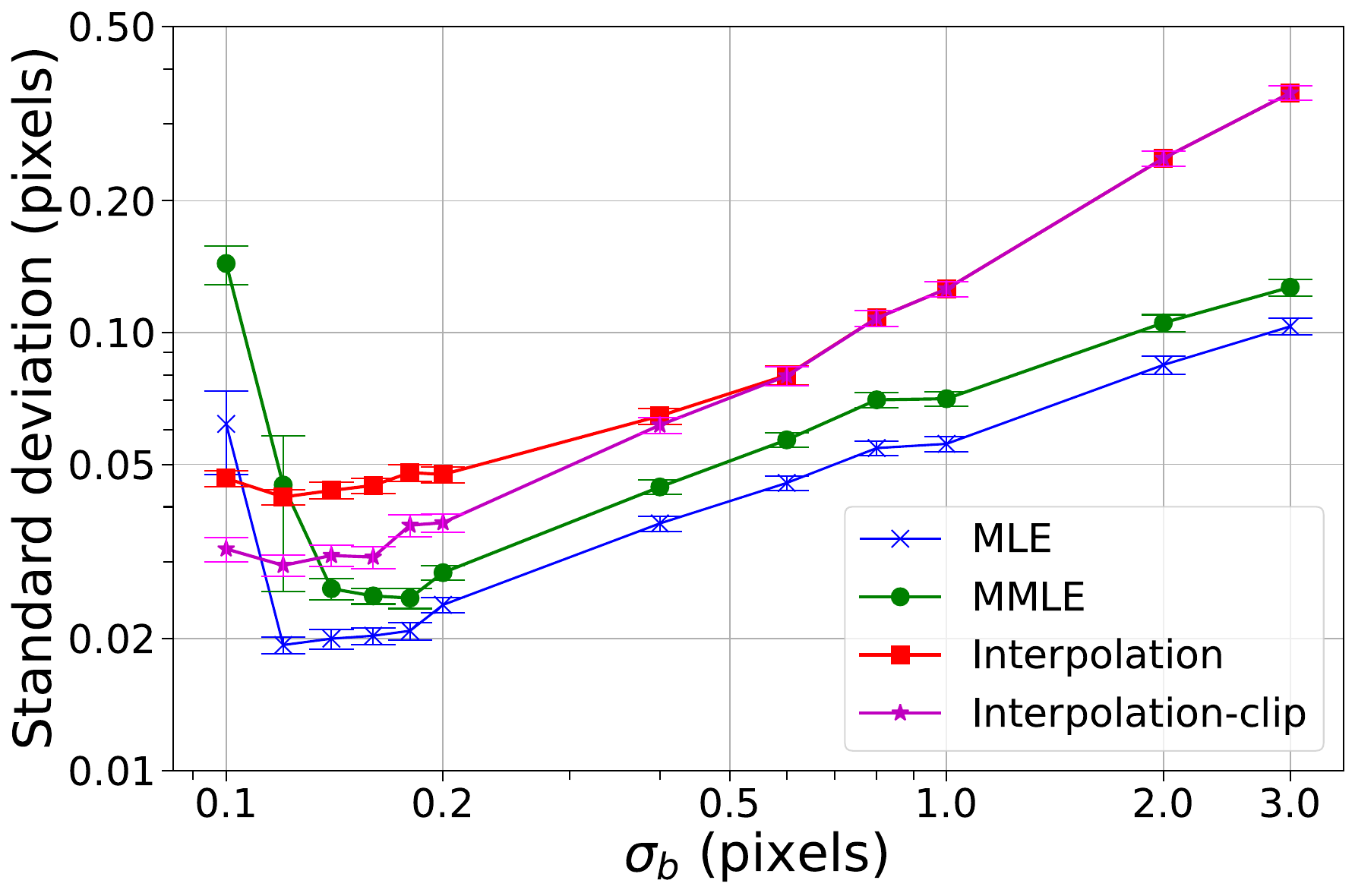} &
    \includegraphics[width=\tableFigureWidth]{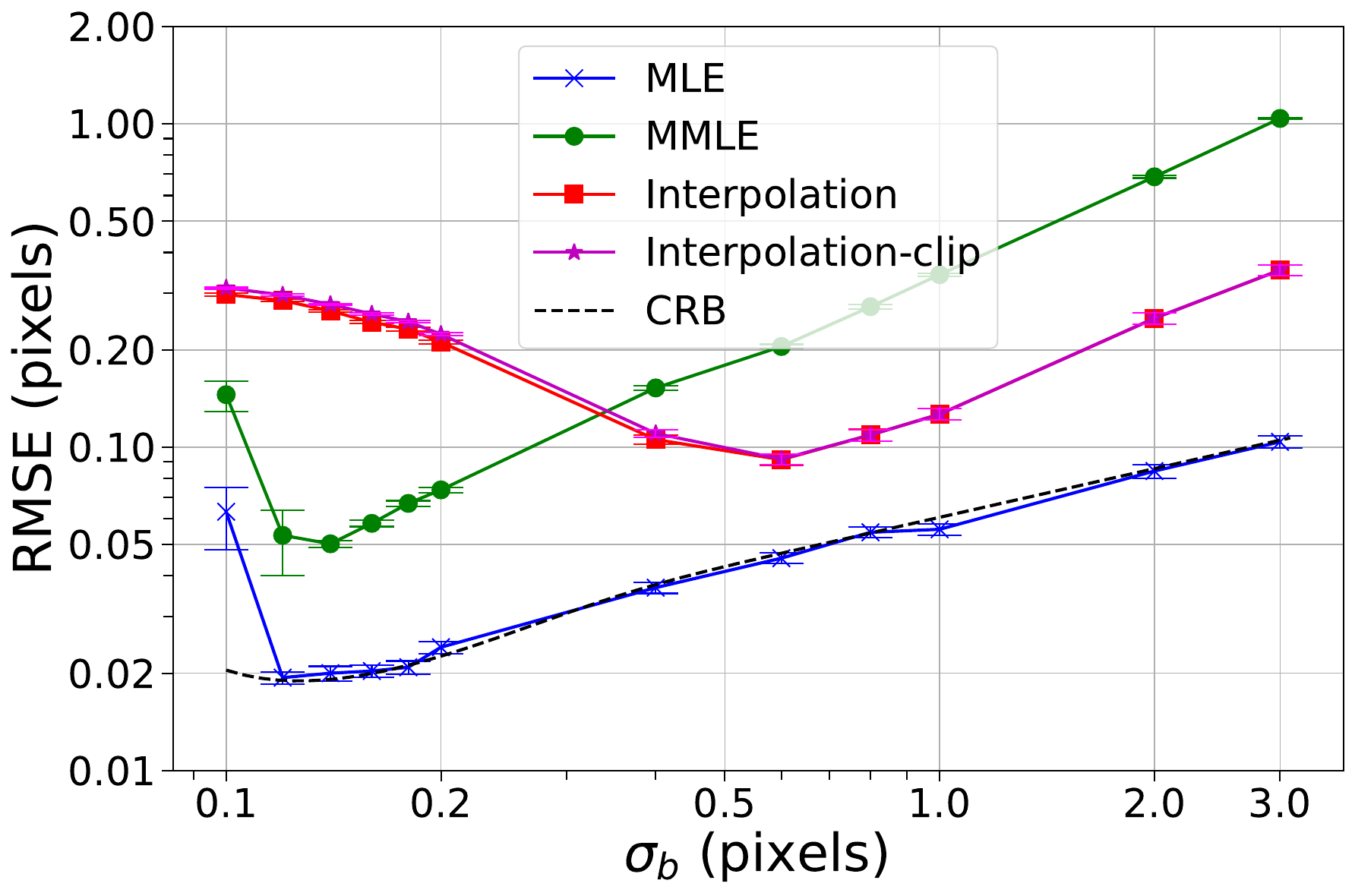} \\[\tightenRow]
    \multicolumn{3}{l}{\footnotesize (b) Varying beam width $\beamSigma$ from 0.1 to 3 with
    edge location $\gamma = 50.2$,
    dose $\lambda=200$,
    and sample SE yields $\eta_1=1$ and $\eta_2=10$.}
    \\[\rowSkip]

    \includegraphics[width=\tableFigureWidth]{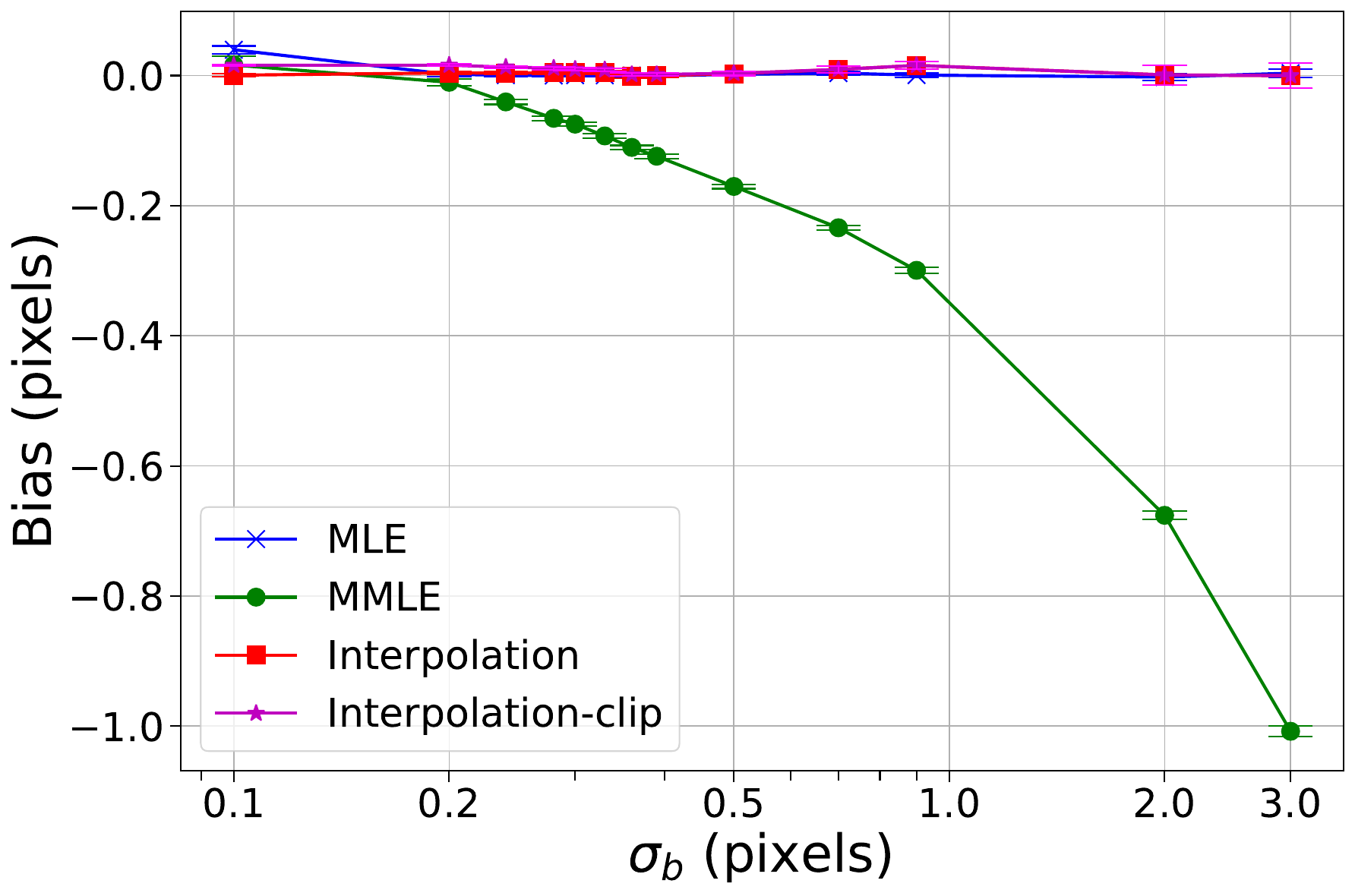} &
    \includegraphics[width=\tableFigureWidth]{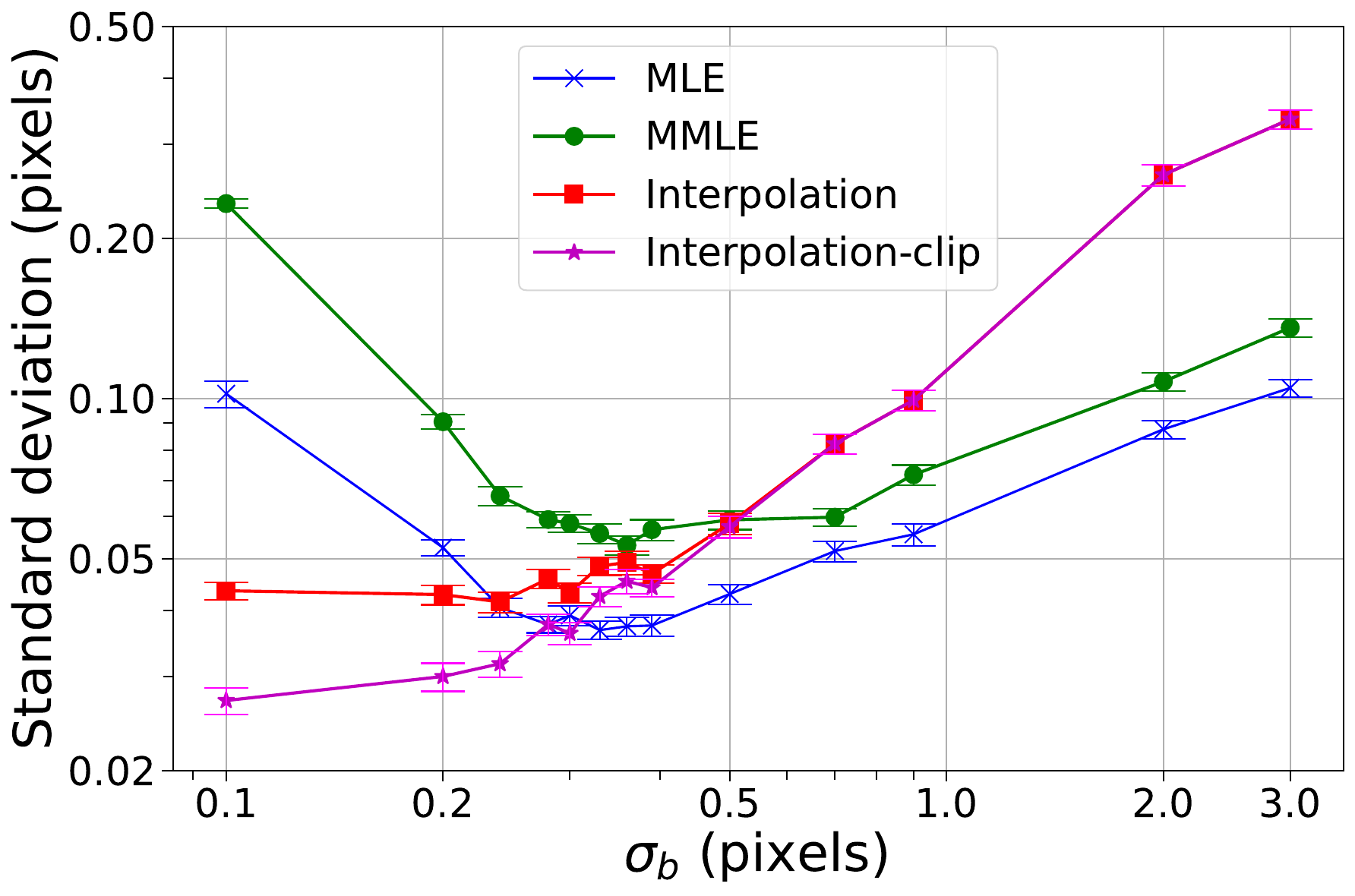} &
    \includegraphics[width=\tableFigureWidth]{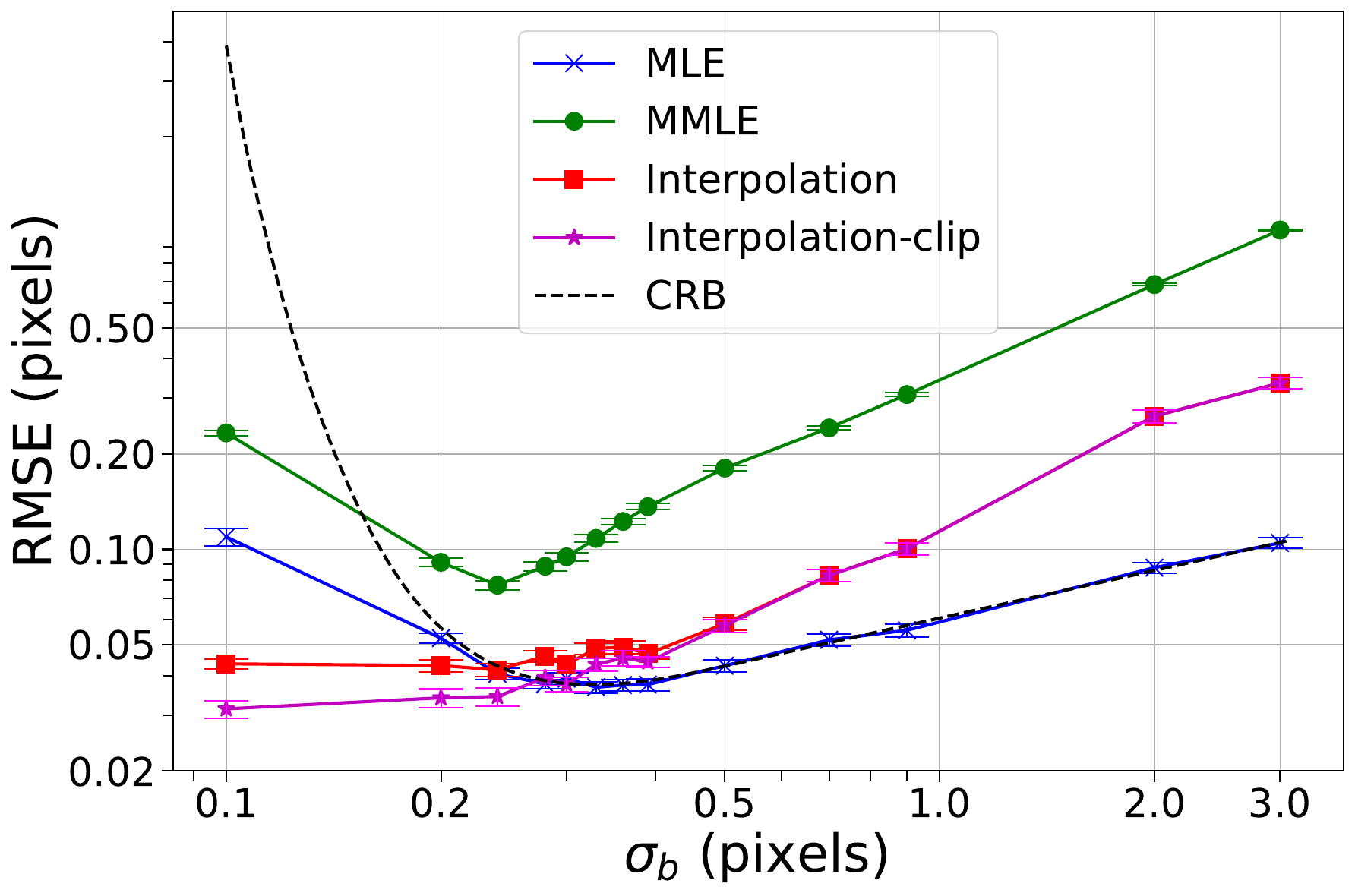} \\[\tightenRow]
    \multicolumn{3}{l}{\footnotesize (c) Varying beam width $\beamSigma$ from 0.1 to 3 with
    edge location $\gamma = 50.5$,
    dose $\lambda=200$,
    and sample SE yields $\eta_1=1$ and $\eta_2=10$.}
    \\[\rowSkip]

    \includegraphics[width=\tableFigureWidth]{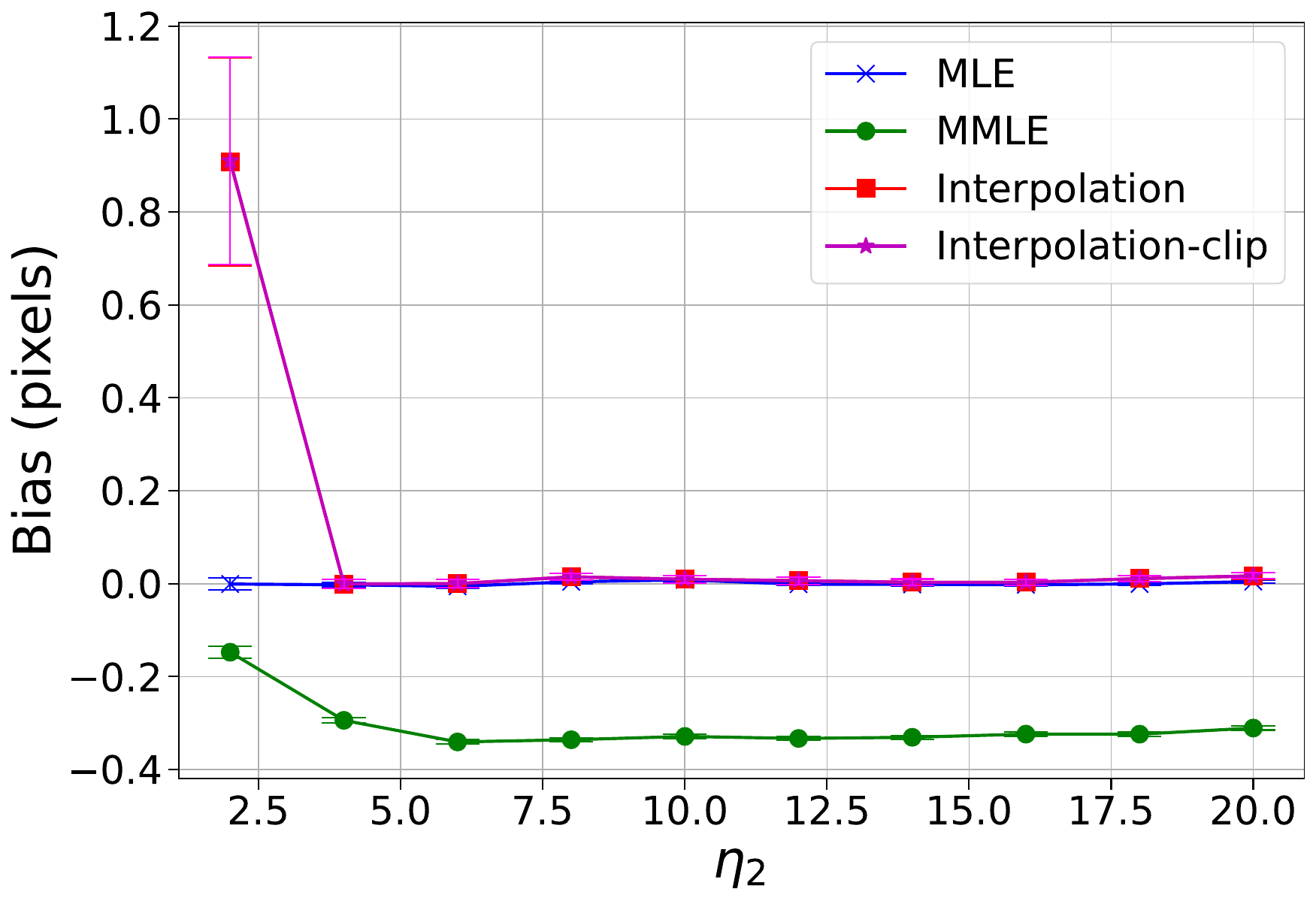} &
    \includegraphics[width=\tableFigureWidth]{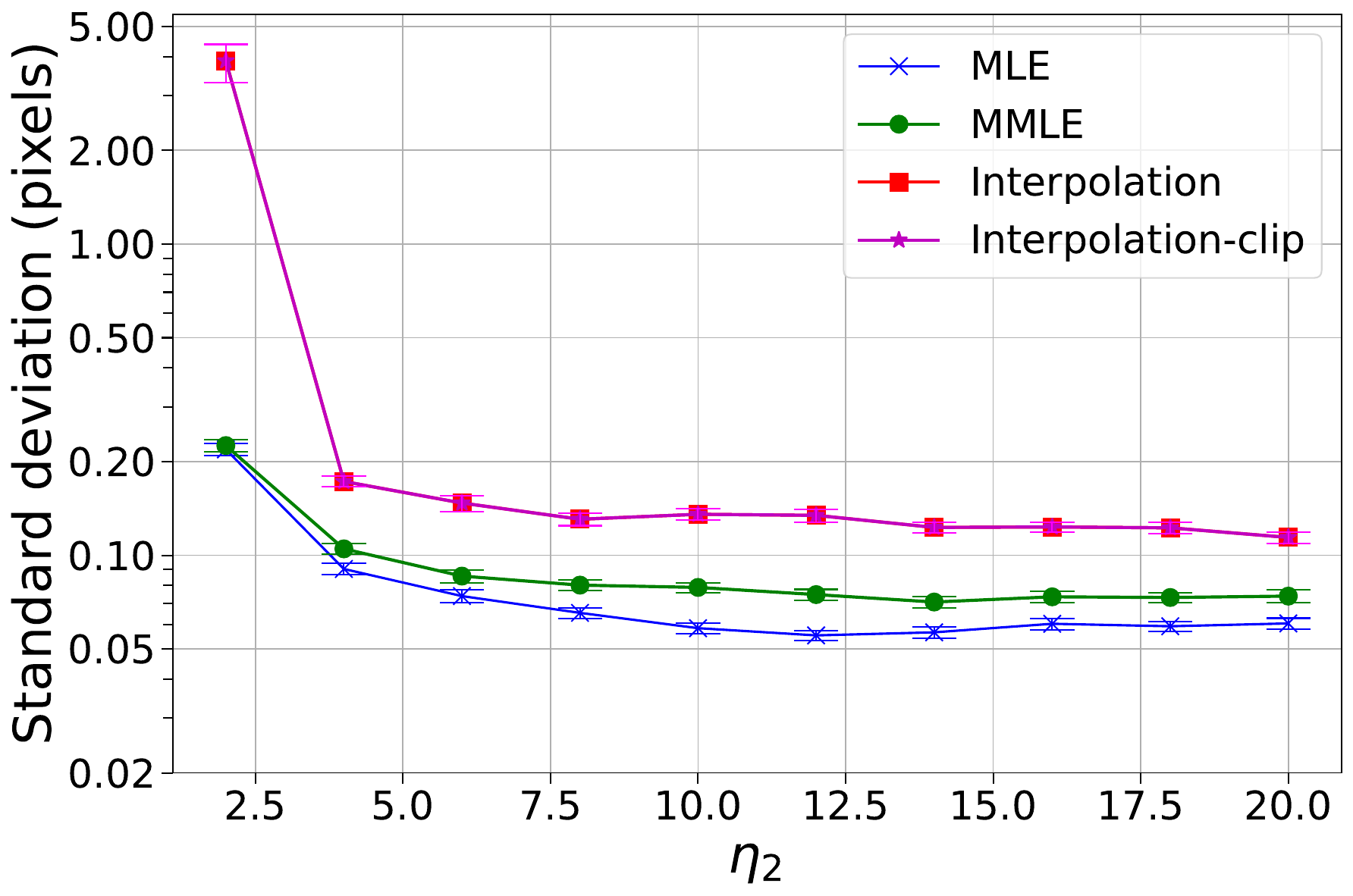} &
    \includegraphics[width=\tableFigureWidth]{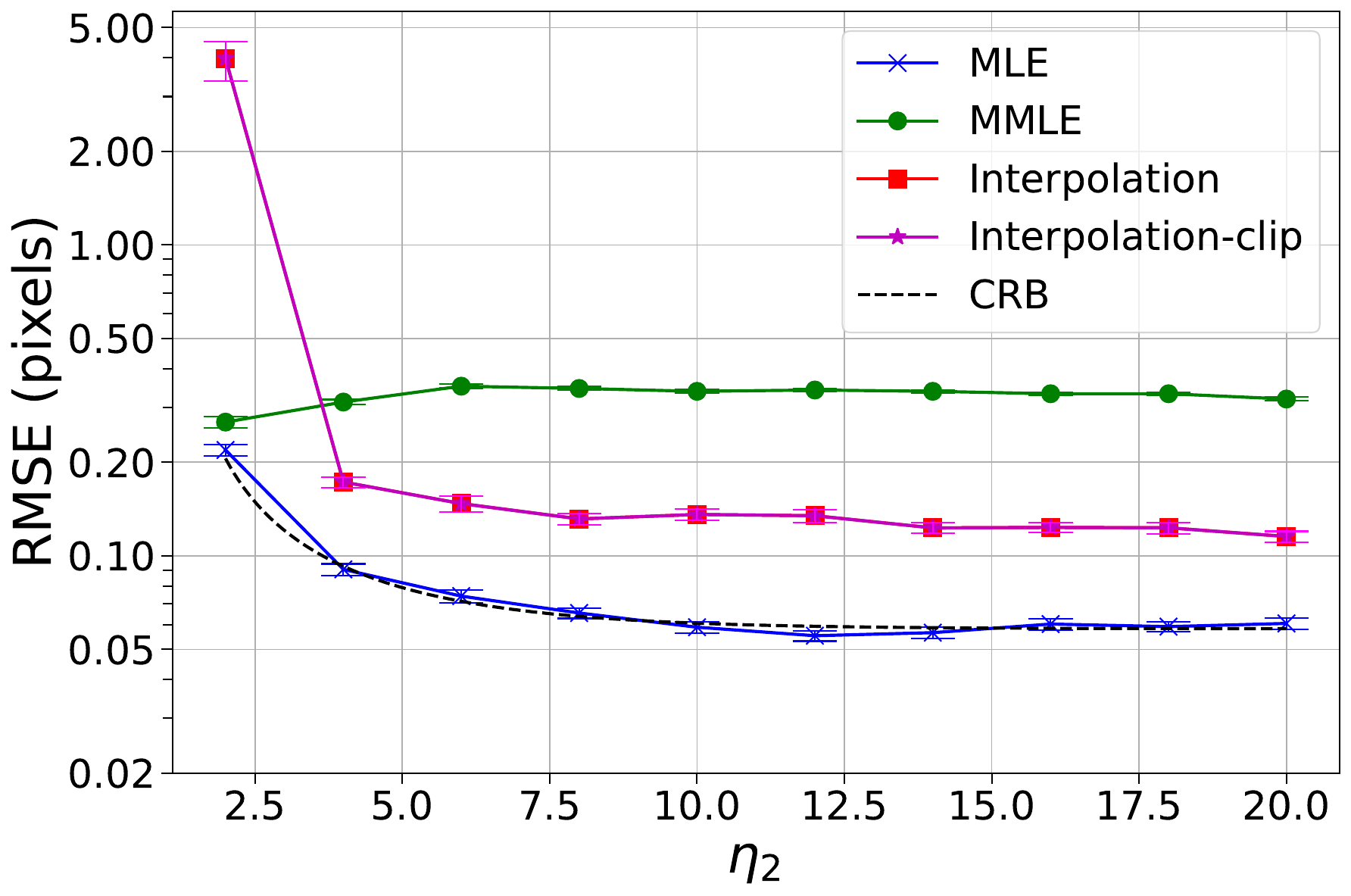} \\[\tightenRow]
    \multicolumn{3}{l}{\footnotesize (d) Varying right side SE yield $\eta_2$ from 2 to 20 with
    edge location $\gamma = 50.2$,
    dose $\lambda=200$,
    beam width $\beamSigma=1$, and
    left side SE yield $\eta_1=1$.}
    
  \end{tabular}
     \caption{Edge localization bias, standard deviation, and RMSE of MLE, MMLE, and interpolation methods for a two-valued sample (see \eqref{eq:two-valued} and \cref{fig:twoValuedSample}).
     For each set of simulation parameters,
     each method is evaluated on a dataset consisting of 300 independent Monte Carlo datasets
     representing a one-dimensional raster scan at locations $\{0,1,\ldots,99\}$.
     The parameters of each experiment are listed below each panel.} 
    \label{subfig:edge_50.2_dose_eta_1_10_RMSE}
\end{figure*}

\subsection{Simulations of Edge Localization}
\label{sec:simresults}
We performed Monte Carlo simulations to evaluate the performances of the four edge estimators,
with the MLE and MMLE implemented by grid search with a grid spacing of 0.01\@.
Each Monte Carlo trial has pseudorandom data generated according to the model in \cref{sec:setup}
on a one-dimensional scan grid of
$\{0,1,\ldots,99\}$
(i.e., $\ell=100$).
Detector noise is not modeled, so a simulated TRM dataset is
$\{ (\mtilde_k,\{\xtilde_{i,k}\}_{i=1}^{\mtilde_k}) \}_{k=0}^{\ell-1}$.
Simulation parameters include the
edge location $\gamma$,
beam width $\beamSigma$,
dose $\lambda$,
and SE yields on each side of the edge, $\eta_1$ and $\eta_2$.
Where applicable, the estimators
use
$\beamSigma$,
$\lambda$,
$\eta_1$, and $\eta_2$; only $\gamma$ is unknown.
In the stylized application considered here,
$\eta_1$ and $\eta_2$ could be estimated well by aggregating data from many pixels far from the edge.
Sets of simulations vary one parameter while keeping others fixed. 
Each reported result is based on 300 Monte Carlo trials,
which allows for computation of (estimated) bias,
(sample) standard deviation, and (estimated) root mean-squared error (RMSE).
The error bars in the bias plots represent $\pm 1$ standard deviation of the sample mean.
We similarly computed $\pm 1$ standard deviation interval endpoints for the sample variance and the sample MSE\@.
The marked error bars in the standard deviation and RMSE plots use
the square roots of these quantities.
The reciprocal of the Fisher information \eqref{eq:FI_gamma} is the {\CR} lower bound on the MSE of any unbiased estimate of $\gamma$ from a TRM dataset.  The square root of this CRB is included in all RMSE plots.

In~\cref{subfig:edge_50.2_dose_eta_1_10_RMSE}(a), we vary $\lambda$ from 20 to 290 while fixing
$\gamma = 50.2$,
$\beamSigma =1$,
$\eta_1 = 1$, and
$\eta_2 = 10$.
The RMSE comparison shows uniform superiority of the MLE over the MMLE and interpolation.

The MLE has very low bias, and its standard deviation and RMSE follow the $\lambda^{-1/2}$ dependence on $\lambda$
predicted by $\Iscan(\gamma \sMid \eta_1,\eta_2,\lambda,\beamSigma)$
in \eqref{eq:FI_gamma}
being linear in $\lambda$.
Along with the $\lambda^{-1/2}$ scaling behavior,
the quantitative match of the RMSE to the square root of the CRB is good here and in most subsequent simulation results.

The MMLE has a negative bias and slightly higher standard deviation than the MLE\@.
The bias being negative can be explained by the
excess variance \eqref{eq:excess-variance-two-valued}
as follows. 
Just to the left of the edge (e.g., at 49 or 50),
the relatively frequent occurrences of
positive outliers in the SE counts,
when interpreted through the convolution model
(Poisson rather than Poisson mixture),
are best explained by the edge having already been reached. 
In other words,
since the Poisson model requires the mean and variance to be equal,
the MMLE incorrectly imputes a larger mixture fraction $q_k$
to account for these outliers. 
Since the mixture fraction values and the edge location are linked by \eqref{eq:mixing-q},
this overestimation of mixture fractions leads to a negative bias in edge estimation.

Here we observe that the interpolation estimators coincide and clipping does not give any advantage.
Both estimators have very low bias at high doses---but significant bias at low doses---and higher standard deviation than the MLE and MMLE\@. 
The bias being positive and vanishing at high enough doses can be explained as follows.
Gross inaccuracy of the interpolation estimator is most likely when there are multiple crossings of the threshold $(\eta_1+\eta_2)/2 = 5.5$.
Far to the left of the edge, $\etaBaseline$ has mean $\eta_1 = 1$ and variance $(\eta_1+\eta_1^2)/\lambda = 2/\lambda$
(see \eqref{eq:E[Y]} and \eqref{eq:var(Y)}).
Thus, even at the lowest shown dose of $\lambda = 20$, spurious threshold crossings (i.e., $\etaBaseline > 5.5$) are very unlikely.
Conversely, far to the right of the edge, $\etaBaseline$ has mean $\eta_2 = 10$ and variance $(\eta_2+\eta_2^2)/\lambda = 110/\lambda$,
and spurious threshold crossings (i.e., $\etaBaseline < 5.5$) happen with nonnegligible probability unless $\lambda$ is large enough.

For panel (b), we choose $\lambda = 200$ so that the dose is high enough to avoid poor performance of interpolation caused by high likelihood of multiple threshold crossings.
We vary $\beamSigma$ from 0.1 to 3 while keeping $\gamma = 50.2$, $\eta_1 = 1$, and $\eta_2 = 10$. 
The RMSE comparison again shows the superiority of the MLE over the MMLE and interpolation.
The signs of biases match the results in panel (a),
with MLE having very low bias,
MMLE having negative bias,
and interpolation having positive bias. 
For MMLE, as $\beamSigma$ shrinks the bias shrinks because the model is substantially mismatched only at raster scan locations where the mixture parameter is close to neither 0 nor 1\@.
Here we first see a performance difference between the interpolation estimators,
with clipping leading to noticeably lower standard deviation values for the lower values of $\beamSigma$.
The biases of the interpolation estimators are nearly equal,
and bias dominates the RMSE at low $\beamSigma$ values, so there is no noticeable difference in the RMSEs
of the interpolation estimators.
For both interpolation estimators, since the dose is relatively large,
the source of bias is different than in panel (a).
As $\beamSigma$ shrinks, the beam interaction with the sample edge vanishes unless the edge is located very close to a raster scan position.
For $\gamma = 50.2$, when $\beamSigma$ is very small and $\lambda$ is sufficiently high, the interpolation estimate should usually be 50.5 because $\etaBaseline \approx 1$ at $k=50$ and $\etaBaseline \approx 10$ at $k=51$, with no spurious threshold crossings.  Thus, the bias is 0.3\@.
This phenomenon is also exhibited by the monotonic increase of the standard deviation of the interpolation estimator as $\beamSigma$ is increased.
We comment on the dependence of the bias on $\gamma$ in discussing panel (c).

In the standard deviation and RMSE plots of panel (b),
we observe the existence of an optimal $\beamSigma$,
as in \cref{fig:fi_gamma_variation}(b).
Here, the Monte Carlo simulations are for a single arbitrary $\gamma$ value whereas \cref{fig:fi_gamma_variation}(b) shows a worst-case over $\gamma$.
Comparing the RMSE of each method at its optimal $\beamSigma$, we see that MLE improves upon interpolation by a factor of 5 and upon MMLE by a factor of 2.5\@.

Panel (c) differs from panel (b) only in changing the edge location to $\gamma = 50.5$.
One striking change is to the bias of the interpolation methods at low values of $\beamSigma$.
This bias is greatly reduced because
$\etaBaseline \approx 1$ at scan position 50
combined with
$\etaBaseline \approx 10$ at scan position 51
gives an interpolation estimate near 50.5
by happenstance;
for interpolation with small $\beamSigma$,
the edge location being at a midpoint between
neighboring scan locations is highly favorable.
In this case, clipping of $\etaBaseline$ gives a slight improvement in standard deviation and RMSE at lower values of $\beamSigma$.
The bias of the MLE is not significantly changed,
but its standard deviation is significantly increased
at low values of $\beamSigma$.
Referring to \cref{fig:fi_gamma_variation}(a),
$\gamma = 50.5$ is a near minimizer of the
FI about $\gamma$ in the scan data.
Indeed, the minima of the RMSE and CRB shown in panel (c)
occur near the $\beamSigma^\star = 0.33$ value found in 
\cref{fig:fi_gamma_variation}(b).

The deviation of MLE performance from the CRB at $\beamSigma = 0.1$ warrants further study.
There is no contradiction in the low exhibited RMSE
because the MLE is not unbiased at this very small $\beamSigma$.
At the two scan locations $k$ closest to the edge (50 and 51),
the normalized distance $|k-\gamma|/\beamSigma$
between the scan location and the edge at $\gamma = 50.5$ is 5,
implying $\min\{q,1-q\} \approx 3 \times 10^{-7}$.
Incident particles are very unlikely to strike on opposite side of the edge relative to the nominal scan location,
so the mixtures are nearly indistinguishable from (single-component) Poisson distributions.

Finally, in panel (d) we vary the right-side SE yield $\eta_2$ from 2 to 20 while fixing
$\eta_1=1$,
$\gamma=50.2$,
$\beamSigma=1$,
and
$\lambda=200$.
As before, the MLE has a very low bias, and its standard deviation and RMSE are the lowest across all values of $\eta_2$.
The root causes of biases of the interpolation estimator
and MMLE are as discussed in regard to panel (a).
Here also we observe that the two interpolation estimators coincide; clipping does not give any advantage.
For both the interpolation estimators, the bias is most pronounced when the gap between $\eta_1$ and $\eta_2$ is small, making spurious threshold crossings relatively more likely.
For MMLE, the bias is smallest where the excess variance of the mixture is smallest.
A trend across most experiments is that the MLE is approximately unbiased and has the lowest standard deviation among all estimators.
It thus has the best performance for varied sample and imaging parameters.
The RMSE in many cases is well below the unit grid spacing, showing the possibility of sub-pixel localization.

\subsection{Experimental Edge Localization}
\label{sec:experimental}
We performed proof-of-concept
experiments
with real data
to test
estimator performances. 
We used a Zeiss Orion Nanolab HIM to image a sample of gold nanofabricated on a silicon substrate.
This sample and the imaging conditions are described in prior work~\cite{Agarwal:2024-PNAS},
from which we use SE yield values of $\eta_1 = 1.82$ for silicon and $\eta_2 = 2.75$ for gold. 
We created three datasets,
each consisting of 51 scan lines with
100 raster scan locations across an edge,
analogous to the Monte Carlo simulations,
with a dose per scan location of $\lambda = 20$.
The dataset generation process is described in more detail in the Supplementary Note~4.

Real HIM datasets have several complexities, such as the conversion of SE counts to detector voltages~\cite{Joy2008} and increase in SE yield at edges~\cite{Mack2015,Mack2016,Mack2017, Villarrubia2015, Cizmar2008, Cazaux2012}.
Furthermore, lack of availability of $\beamSigma$ and the ground truth edge location make direct application and evaluation of the methods described here challenging.
We converted detector voltages to SE counts using a hard thresholding based on prior work~\cite{Agarwal:2024-PNAS},
and we used a defocused beam to reduce edge brightening.
Since $\beamSigma$ is not otherwise available,
we computed a joint MLE
from all the available data (all lines of the three datasets)
of the beam width and
collection of
edge locations. 
To generate proxies for the ground truth edge location $\gamma$,
we applied the interpolation estimator to an emulation of the very high dose of $\lambda=1020$
by concatenating the data from 51 line scans.
As we observe in \cref{subfig:edge_50.2_dose_eta_1_10_RMSE}(a), the interpolation estimate has very low bias and variance as the dose increases, making it a good ground truth proxy. 
 
The results of running the estimators on experimental data are shown in \cref{table:experimental_results}.
We observed that both the interpolation methods gave the same performance because clipping never influenced a $\gamma$ estimate.
Hence these estimators are not shown separately in the table.
The MLE has lowest bias across all the three datasets,
followed by the MMLE and interpolation methods.
Both MLE and MMLE have lower standard deviations than the interpolation methods,
similar to the simulation results. 
MLE consistently achieves the lowest RMSE across all three datasets,
outperforming the interpolation method by a factor of 4 to 7 and surpassing MMLE by about 1.5 times.
These results
align with the trends observed in our simulation experiments.

\begin{table}
\caption{Experimental results with three HIM datasets}
\centering
\scalebox{0.86}{
\begin{tabular}{*{3}{|*{3}{r}}|}
\hline
\multicolumn{3}{|c|}{Bias} 
& \multicolumn{3}{c|}{Standard deviation} 
& \multicolumn{3}{c|}{RMSE} \\
\hline
\multicolumn{1}{|c}{Interp}
& \multicolumn{1}{c}{MMLE}
& \multicolumn{1}{c|}{MLE}
& \multicolumn{1}{c}{Interp}
& \multicolumn{1}{c}{MMLE}
& \multicolumn{1}{c|}{MLE}
& \multicolumn{1}{c}{Interp}
& \multicolumn{1}{c}{MMLE}
& \multicolumn{1}{c|}{MLE} \\
\hline
 16.8 & -3.3 & -1.0 & 5.4 & 2.4 & 2.3 & 17.6 & 4.1 & \textbf{2.5} \\
 -11.7 & -4.9 & -2.9 & 7.4 & 1.8 & 1.8 & 13.9 & 5.3 & \textbf{3.4} \\
 -7.8 & -2.0 & -0.1 & 6.6 & 1.9 & 2.0 & 10.3 & 2.7 & \textbf{2.0} \\
\hline
\end{tabular}
}
\\[2ex]
\label{table:experimental_results}
\end{table}

\section{Conclusion}
Where secondary electron imaging is quantitative,
the goal is generally to estimate SE yield at each raster scan location
to create a micrograph.
Any inferences to be made from SE imaging are then further computations with the micrograph as input,
without regard to the signals involved in producing the micrograph.
The introduction of time-resolved measurement to SE imaging in~\cite{PengMBBG:20, PengMBG:21, AgarwalPG:23, Agarwal:2024-PNAS}
created the potential to estimate the SE count distribution rather than only its mean.
This paper is a first attempt to exploit the estimation of an SE count distribution at each raster scan location to achieve a metrology task.

The paper considers a simple, illustrative model for edge localization 
in which the sample emits Poisson-distributed numbers of SEs with different means
for particles incident on each side of an edge.
Nonzero beam width then creates a two-component Poisson mixture for
the observed SE count.
Notably, the mean of this distribution is described by a convolution,
and such convolutional models are common in the literature.
We show, however, that exploiting the mixture model enables more accurate edge localization.

Expanding upon this work could lead to valuable methods for semiconductor metrology,
especially critical dimension measurements and line width roughness analyses.
However, several complicating factors must be overcome.
The increase of SE yield around edges---\emph{edge brightening}---changes the SE yield variation from the two-valued functions considered here, thereby altering the distribution of SE counts.
Several works explain physical bases for edge brightening and give mathematical models to approximate the SE signal profile across an edge~\cite{Mack2015,Mack2016,Mack2017}.
These have been used to treat high-resolution reconstruction as a deconvolution problem.
In conjunction with these works, our mixture modeling framework may further improve localization accuracy.

Beyond the transverse spatial modeling that is central to the demonstrated edge localization in this paper,
a more expansive view is to
recognize that distinct incident particles may interact with the sample
in different ways and thus generate mixture SE count distributions.
This may hold potential for inferring material properties of the sample.
Current techniques identify surface properties,
such as grain size and grain orientation using SE yield variation at fine spatial resolution~\cite{Turner2003}.
Improved modeling of SE count distributions may give more direct manifestations of
alloy compositions and electronic states.

\newcommand{\SortNoop}[1]{}

\newpage

\onecolumn

\setcounter{page}{1}

\begin{center}
    \huge Supplementary Material
\end{center}

\section*{Note 1: Derivation of Fisher information about $(q,\eta_1,\eta_2)$ in $X$}

In \cref{subsec:FI_deterministic},
we assert that the {\CR} bound for the estimation of $\eta$ from mixture observations $X_1,\,X_2,\ldots,\,X_m$
when $q$, $\eta_1$ and $\eta_2$ are unknown is numerically indistinguishable from the MSE of
the sample mean estimator.
Here we present the basis for this claim.
We first derive the Fisher information about unknown parameter vector $\Theta = (q,\eta_1,\eta_2)$,
and we then use a reparameterization to yield the CRB on unbiased estimates of $\eta$.

For $(j,k) \in \{1,2,3\}^2$, the $(j,k)$ element of $\Imixture_X(\Theta)$ is given by
\begin{equation}
    \label{eq:Fisher-def}
    [\Imixture_{X}(\Theta)]_{j,k} = \E{\left(
    \pdv{\log \mathrm{P}_X(X \sMid \Theta)}{\Theta_j}
    \right)
    \left(
    \pdv{\log \mathrm{P}_X(X \sMid \Theta)}{\Theta_k}
    \right)} ,
\end{equation}
where the PMF of $X$ is given by \eqref{eq:mixpoissondistr}.
The three required derivatives are
\begin{subequations}
\label{eq:FI_diff_all}
\begin{align}
    \pdv{\log \mathrm{P}_X(x \sMid \Theta)}{\Theta_1}
      &= \pdv{\log \mathrm{P}_X(x \sMid \Theta)}{q}
       = \frac{\eta_2^x e^{-\eta_2}-\eta_1^x e^{-\eta_1}}
              {x! \, \mathrm{P}_X(x \sMid \Theta)},
              \label{eq:FI_diff_1} \\
    \pdv{\log \mathrm{P}_X(x \sMid \Theta)}{\Theta_2}
      &= \pdv{\log \mathrm{P}_X(x \sMid \Theta)}{\eta_1}
        = 
         \frac{(1-q) (x - \eta_1 ) \eta_1^{x-1} e^{-\eta_1}}
              {x! \, \mathrm{P}_X(x \sMid \Theta)},
              \label{eq:FI_diff_2} \\
    \pdv{\log \mathrm{P}_X(x \sMid \Theta)}{\Theta_3}
      &= \pdv{\log \mathrm{P}_X(x \sMid \Theta)}{\eta_2}
       =
         \frac{q (x - \eta_2 ) \eta_2^{x-1} e^{-\eta_2}}
              {x! \, \mathrm{P}_X(x \sMid \Theta)},
              \label{eq:FI_diff_3} 
\end{align}
\end{subequations}
where \eqref{eq:FI_diff_1} matches \eqref{eq:dlog_Px_dq}.
Based on these derivatives, $\Imixture_X(\Theta)$ has six distinct elements:
\begin{subequations}
\label{eq:Imixture_x_Theta}
\begin{align}
    [\Imixture_{X}(\Theta)]_{1,1} &= \E{
    \frac{
       (\eta_2^x e^{-\eta_2}-\eta_1^x e^{-\eta_1})^2
    }{
       \left( (1-q) \eta_1^x e^{-\eta_1} + q \eta_2^x e^{-\eta_2} \right)^2
    }
    }
    \nonumber \\
    &=
    \sum_{x=0}^\infty
    \frac{
       (\eta_2^x e^{-\eta_2}-\eta_1^x e^{-\eta_1})^2
    }{
       x! \left( (1-q) \eta_1^x e^{-\eta_1} + q \eta_2^x e^{-\eta_2} \right)
    }
    ,
    \\
    [\Imixture_{X}(\Theta)]_{1,2} &= \E{
    \frac{
    (\eta_2^x e^{-\eta_2}-\eta_1^x e^{-\eta_1})
    ((1-q)(x-\eta_1)\eta_1^{x-1}e^{-\eta_1})
    }{
        \left( (1-q) \eta_1^x e^{-\eta_1} + q \eta_2^x e^{-\eta_2} \right)^2
    }}   
    \nonumber \\
    &=
    \sum_{x=0}^\infty
    \frac{
    (\eta_2^x e^{-\eta_2}-\eta_1^x e^{-\eta_1})
    ((1-q)(x-\eta_1)\eta_1^{x-1}e^{-\eta_1})
    }{
     x! \left( (1-q) \eta_1^x e^{-\eta_1} + q \eta_2^x e^{-\eta_2} \right)
    }
    ,
    \\
    [\Imixture_{X}(\Theta)]_{1,3} &= \E{
    \frac{
    (\eta_2^x e^{-\eta_2}-\eta_1^x e^{-\eta_1})
    (q (x - \eta_2 ) \eta_2^{x-1} e^{-\eta_2})
    }{
    \left( (1-q) \eta_1^x e^{-\eta_1} + q \eta_2^x e^{-\eta_2} \right)^2
    }
    }
    \nonumber \\
    &=
    \sum_{x=0}^\infty
    \frac{
    (\eta_2^x e^{-\eta_2}-\eta_1^x e^{-\eta_1})
    (q (x - \eta_2 ) \eta_2^{x-1} e^{-\eta_2})
    }{
    x! \left( (1-q) \eta_1^x e^{-\eta_1} + q \eta_2^x e^{-\eta_2} \right)
    }
    , 
    \\
    [\Imixture_{X}(\Theta)]_{2,2} &=\E{
    \frac{
    ((1-q) (x - \eta_1 ) \eta_1^{x-1} e^{-\eta_1})^2
    }{
     \left( (1-q) \eta_1^x e^{-\eta_1} + q \eta_2^x e^{-\eta_2} \right)^2
    }
    }
    \nonumber \\
    &=   
     \sum_{x=0}^\infty
     \frac{
     ((1-q) (x - \eta_1 ) \eta_1^{x-1} e^{-\eta_1})^2
     }{
      x! \left( (1-q) \eta_1^x e^{-\eta_1} + q \eta_2^x e^{-\eta_2} \right)
     }
     ,
     \\
     [\Imixture_{X}(\Theta)]_{2,3} &=\E{
     \frac{
     ((1-q) (x - \eta_1 ) \eta_1^{x-1} e^{-\eta_1})
     (q (x - \eta_2 ) \eta_2^{x-1} e^{-\eta_2})
     }{
    \left( (1-q) \eta_1^x e^{-\eta_1} + q \eta_2^x e^{-\eta_2} \right)^2
     }
     }
     \nonumber \\
     &=
     \sum_{x=0}^\infty
     \frac{
     ((1-q) (x - \eta_1 ) \eta_1^{x-1} e^{-\eta_1})
     (q (x - \eta_2 ) \eta_2^{x-1} e^{-\eta_2})
     }{
    x! \left( (1-q) \eta_1^x e^{-\eta_1} + q \eta_2^x e^{-\eta_2} \right)
     },
     \\
     [\Imixture_{X}(\Theta)]_{3,3}&= \E{
     \frac{
     (q (x - \eta_2 ) \eta_2^{x-1} e^{-\eta_2})^2
     }{
     \left( (1-q) \eta_1^x e^{-\eta_1} + q \eta_2^x e^{-\eta_2} \right)^2
     }
     }
     \nonumber \\
     &=
      \sum_{x=0}^\infty
      \frac{
      (q (x - \eta_2 ) \eta_2^{x-1} e^{-\eta_2})^2
      }{
       x! \left( (1-q) \eta_1^x e^{-\eta_1} + q \eta_2^x e^{-\eta_2} \right)
      }
     ,
\end{align}
\end{subequations}
with the remaining three given by symmetry.

The parameter $\eta$ is determined by $\Theta$ according to
\begin{equation}
    \eta = (1-q)\eta_1 + q \eta_2 = (1-\theta_1)\theta_2 + \theta_1 \theta_3.
\end{equation}
Thus, we have the vector of partial derivatives
\begin{equation}
    \label{eq:J}
    J = \begin{bmatrix}
\Frac{\partial \eta}{\partial \theta_1} \\
\Frac{\partial \eta}{\partial \theta_2} \\
\Frac{\partial \eta}{\partial \theta_3} 
\end{bmatrix}
=
\begin{bmatrix}
\theta_3 - \theta_2 \\
1 - \theta_1 \\
\theta_1 
\end{bmatrix}
=
\begin{bmatrix}
\eta_2 - \eta_1 \\
1 - q \\
q 
\end{bmatrix}.
\end{equation}
Using the reparametrization for FI with a vector of parameters~\cite[Ch.~2 (6.16)]{LehmannC:98},
the CRB for the variance of an unbiased estimate of $\eta$ is
\begin{equation}
\label{eq:transformed_IMixture_inverse}
    J^T (\Imixture_X(\Theta))^{-1} J.
\end{equation}
We are unable to analytically simplify
\eqref{eq:transformed_IMixture_inverse}
after substitution of
\eqref{eq:Imixture_x_Theta} and
\eqref{eq:J}.
Numerically,
\eqref{eq:transformed_IMixture_inverse}
indistinguishable from 
$(1-q)\eta_1 + q\eta_2 + q(1-q)(\eta_2-\eta_1)^2$,
which is the variance of the sample mean estimator for $m=1$.

\newpage
\section*{Note 2: Normalized Fisher information limits of $Y$}
\label{app:NFI-limits}

\subsection*{Low-dose limit}
To evaluate $\lim_{\lambda \rightarrow 0} \mathcal{I}_Y(q \sMid \eta_1, \eta_2, \lambda)/\lambda$,
we first find $\lambda \rightarrow 0$ approximations of 
$\mathrm{P}_Y(y)$ in \eqref{eq:PY}
and
$\numer(y)$ in \eqref{eq:numer}
for substitution into \eqref{eq:I_Y_q_sum}.
Because of the form of \eqref{eq:I_Y_q_sum},
having both $\numer(y)$ and $\mathrm{P}_Y(y)$
up to linear in $\lambda$ will give
$\mathcal{I}_Y(q \sMid \eta_1,\eta_2,\lambda)$
up to the linear term in $\lambda$,
which is what we desire for the limit of FI normalized by $\lambda$.

To approximate $\mathrm{P}_Y(y)$ up to the linear term in $\lambda$,
we can truncate the outer summation in \eqref{eq:PY} at $m=1$.
Using
\begin{equation}
  \label{eq:binProb_low_lambda}
   \binProb_{0,0} = 1,
   \qquad
   \binProb_{1,0} = 1-q,
   \qquad
   \binProb_{1,1} = q,
\end{equation}
and
\begin{equation}
  \label{eq:mixMean_low_lambda}
   \mixMean_{0,0} = 0,
   \qquad
   \mixMean_{1,0} = \eta_1,
   \qquad
   \mixMean_{1,1} = \eta_2,
\end{equation}
we obtain
\begin{subequations}
    \label{eq:PY_low_lambda}
\begin{equation}
    \label{eq:PY_0_low_lambda}
    \mathrm{P}_Y(0) \approx e^{-\lambda} + \lambda e^{-\lambda}
    \left[ (1-q)e^{-\eta_1} + q e^{-\eta_2} \right]
\end{equation}
and, for $y \geq 1$,
\begin{equation}
  \label{eq:PY_y_low_lambda}
    \mathrm{P}_Y(y) \approx \lambda e^{-\lambda}
    \left[ (1-q) \frac{\eta_1^y}{y!} e^{-\eta_1}
            + q \frac{\eta_2^y}{y!} e^{-\eta_2} \right].
\end{equation}
\end{subequations}

Similarly,
to approximate $\numer(y)$ up to the linear term in $\lambda$,
we can truncate the outer summation in \eqref{eq:numer} at $m=1$.
Using
\begin{equation}
  \label{eq:binProbDiff_low_lambda}
   \binProbDiff_{0,0} = 0,
   \qquad
   \binProbDiff_{1,0} = -1,
   \qquad
   \binProbDiff_{1,1} = 1,
\end{equation}
and \eqref{eq:mixMean_low_lambda},
we obtain
\begin{equation}
    \label{eq:numer_y_low_lambda}
    \numer(y) \approx
          \lambda e^{-\lambda}
    \left[- \frac{\eta_1^y}{y!} e^{-\eta_1}
          + \frac{\eta_2^y}{y!} e^{-\eta_2} \right].
\end{equation}
Now
\begin{subequations}
  \label{eq:I_Y_low_lambda_summand}
\begin{equation}
  \lim_{\lambda \rightarrow 0}
    \frac{1}{\lambda}
    \frac{(L(0))^2}{\mathrm{P}_Y(0)}
    = 0
\end{equation}
and, for $y \geq 1$,
\begin{equation}
  \lim_{\lambda \rightarrow 0}
    \frac{1}{\lambda}
    \frac{(L(y))^2}{\mathrm{P}_Y(y)}
    = 
    \frac{1}{y!}
    \frac{
    \left(  \eta_1^y e^{-\eta_1}
          - \eta_2^y e^{-\eta_2} \right)^2
    }
    {
           (1-q) \eta_1^y e^{-\eta_1}
            + q  \eta_2^y e^{-\eta_2} 
    } .
\end{equation}
\end{subequations}
Substituting \eqref{eq:I_Y_low_lambda_summand}
into \eqref{eq:I_Y_q_sum}
yields our final result of
\begin{equation}
  \lim_{\lambda \rightarrow 0}
    \frac{\mathcal{I}_Y(q \sMid \eta_1, \eta_2, \lambda)}{\lambda}
    = 
    \sum_{y=1}^\infty
    \frac{1}{y!}
    \frac{
    \left(  \eta_1^y e^{-\eta_1}
          - \eta_2^y e^{-\eta_2} \right)^2
    }
    {
           (1-q) \eta_1^y e^{-\eta_1}
            + q  \eta_2^y e^{-\eta_2} 
    } .
\end{equation}

\subsection*{High-dose limit}

We will obtain a high-dose limit based on a Gaussian approximation, so let us first state the general property of FI in a Gaussian random variable that we will use.
Let $S \sim \mathcal{N}(\mu(q),v(q))$.
Then
\begin{equation}
    \label{eq:Gaussian-FI}
    \mathcal{I}_S(q) = \frac{\mu'(q)^2}{v(q)} + \frac{[v'(q)]^2}{2v(q)^2}.
\end{equation}
When $\mu(q) = q$ and $v(q)$ is constant, this reduces to the familiar reciprocal of the variance.

For large dose $\lambda$, the conventional observation $Y$ is likely to be a sum of a large number of independent and identically distributed random variables.
Hence, we expect a Gaussian approximation for $Y/\lambda$ to hold.
Indeed,
Theorem~4 in Section~VIII.4 of
W.~Feller, \emph{An Introduction to Probability Theory and Its Applications}, 2nd~ed., John Wiley \& Sons, 1971, vol.~2,
gives a central limit theorem for random sums that applies in our setting.
Specifically, 
by rescaling of \eqref{eq:E[Y]} and \eqref{eq:var(Y)},
$Y/\lambda$ is well approximated by Gaussian $S$ with
\begin{align}
    \mu(q)  = \eta = (1-q)\eta_1 + q\eta_2, 
    \qquad
      v(q)  = \left(\eta + \eta^2 + q(1-q)(\eta_2-\eta_1)^2 \right)/\lambda.
\end{align}
We thus have
\begin{align}
    \mu'(q)  = \eta_2 - \eta_1, 
    \qquad
      v'(q)  = \left((\eta_2-\eta_1) + 2\eta(\eta_2-\eta_1) + (1-2q)(\eta_2-\eta_1)^2 \right)/\lambda.
\end{align}
Substituting these in \eqref{eq:Gaussian-FI} makes the first term proportional to $\lambda$ and the second term a constant with respect to $\lambda$.
The second term is therefore negligible when taking a $\lambda \rightarrow \infty$ limit.
We thus obtain
\begin{equation}
    \lim_{\lambda\rightarrow\infty} \frac{\mathcal{I}_Y(q \sMid \eta_1, \eta_2, \lambda)}{\lambda}
     = \frac{(\eta_2-\eta_1)^2}{\eta + \eta^2 + q(1-q)(\eta_2-\eta_1)^2}.
\end{equation}

\newpage
\section*{Note 3: Maximin optimal beam width}

\begin{figure}[h]
 \centering
  \begin{tabular}{c}
    \includegraphics[width=2.1\singleColumnGraphWidth]{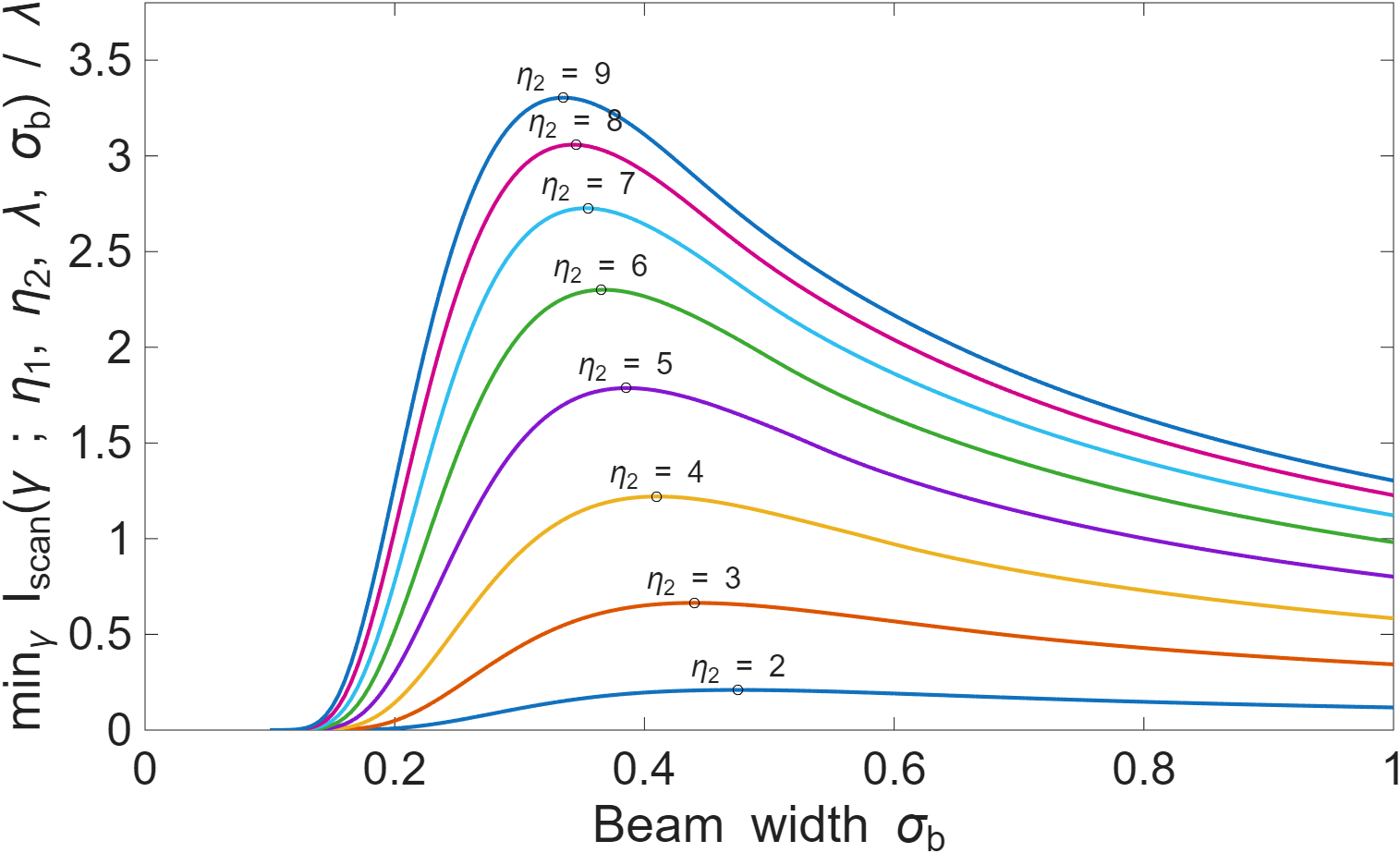} \\
    {\footnotesize (a) $\eta_1 = 1$} \\[4ex]
    \includegraphics[width=2.1\singleColumnGraphWidth]{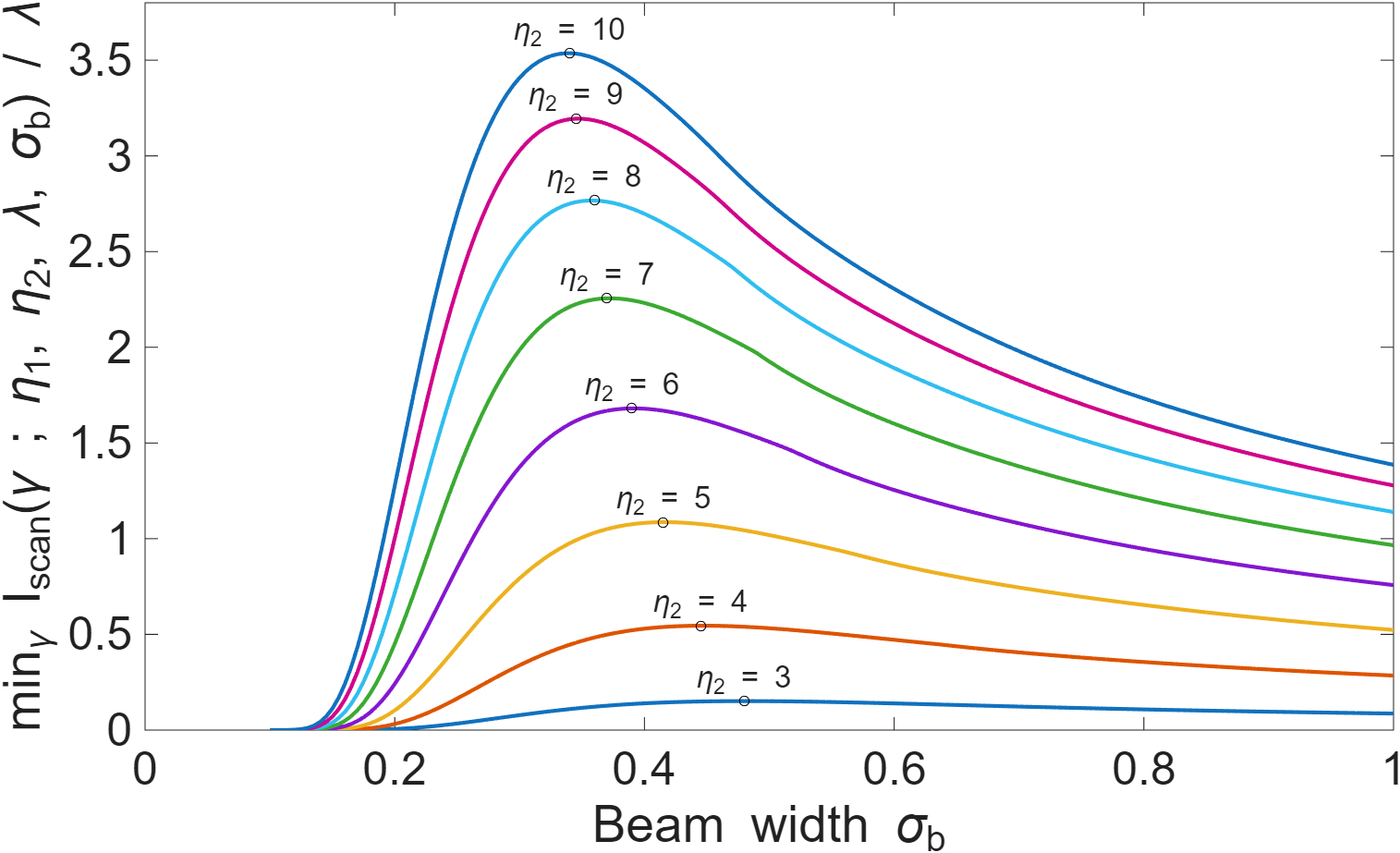} \\
    {\footnotesize (b) $\eta_1 = 2$} \\
  \end{tabular}
 \caption{Study of the mild dependence of $\beamSigma^\star$ on $(\eta_1,\eta_2)$.
   The worst case (minimum) over $\gamma$ of the Fisher information about the edge location $\gamma$ in a full scan at $\{0,\,1,\,\ldots,\,99\}$ is plotted as a function of beam width $\beamSigma$.
   The $\beamSigma$ that maximizes this FI is the maximin optimal value for the beam width.
   Curves are plotted here for (a) $\eta_1 = 1$ and (b) $\eta_1 = 2$ and various values of $\eta_2$.}
\label{fig:beamSigmaOpt_variation}
\end{figure}

\newpage
\section*{Note 4: Details of experimental setup and simplifying assumptions for data processing}

We used a Zeiss Orion Nanolab HIM, operated at 30\,keV beam energy and 0.1\,pA beam current.
The beam was scanned over a sample consisting of 1-micron squares of gold nanofabricated on a silicon substrate. 
The pixel dwell time was 32\,microseconds,
which corresponds to a dose of 20 ions per pixel. 
The sample was used in \cite{Agarwal:2024-PNAS};
see \cite[Fig.~3]{Agarwal:2024-PNAS}, which uses data from the same imaging session.

As discussed in \cref{sec:experimental}, the operation of real HIMs includes several additional complexities that make it difficult to directly apply the estimators discussed in the paper. 
The SE detector does not directly count SEs; instead, it produces voltage pulses whose mean heights are proportional to the number of SEs that produce that pulse. 
In prior work we modeled the distribution of this voltage as Gaussian and measured its mean and variance~\cite{Agarwal:2024-PNAS}.
For this paper, we implemented a hard-thresholding step to convert the measured SE pulse voltages into SE counts.
This hard thresholding was accomplished dividing each pulse voltage by the mean pulse voltage for 1 SE (measured to be 0.163 V in~\cite{Agarwal:2024-PNAS})
and rounding to the nearest integer. 

Further, the SE yield increases over its intrinsic value near sharp edges due to the greater surface area over which SEs can escape, causing a well-known `edge brightening'~\cite{Mack2015, Villarrubia2015}.
Accurate modelling of this effect would require a continuous-valued mixture.
Therefore, a complete probabilistic model for image formation would include the distribution of SE detector voltages as well as two-dimensional distributions for the beam profile and resulting mixture distributions.
To minimize edge brightening, we intentionally defocused the ion beam so that it was significantly wider than the gap between consecutive scan positions. 
These simplifying procedures preclude a true demonstration of improved high-resolution imaging or a quantitative match with simulations, but, as described in the paper, the edge estimates we obtained match the theoretical trends of bias and standard deviation seen in Monte Carlo simulations.

\end{document}